\documentclass[aps,pra,reprint,superscriptaddress,longbibliography,nofootinbib,noeprint]{revtex4-2}

\usepackage[english]{babel} 
\usepackage[utf8]{inputenc}
\usepackage[T1]{fontenc}
\usepackage{amsmath}
\usepackage{amssymb}
\usepackage{amsthm}
\usepackage{mathtools}
\usepackage{bbold,bbm}
\usepackage{graphicx}
\usepackage{verbatim}
\usepackage{physics}
\usepackage{listings}
\usepackage{float}
\usepackage{floatflt}
\usepackage{blindtext}

\usepackage{bm,graphicx}
\usepackage[utf8]{inputenc}
\usepackage[english]{babel}
\usepackage[dvipsnames]{xcolor}
\graphicspath{{./Figures/}}

\usepackage[caption=false]{subfig}
\usepackage{tikz}\usetikzlibrary{decorations.markings,decorations.pathreplacing,decorations.pathmorphing,arrows,arrows.meta,calc,shapes.arrows,shapes.geometric,quantikz2}

\usepackage{siunitx}
\usepackage[colorlinks, linkcolor=red,citecolor=blue,urlcolor=blue]{hyperref}
\usepackage[all]{hypcap}

\begin{document}

\title{Quantum sensing from gravity as a universal dephasing channel for qubits}

\author{Alexander V. Balatsky}
\affiliation{Nordita, Stockholm University and KTH Royal Institute of Technology, Hannes Alfv\'ens v\"ag 12, SE-106~91 Stockholm, Sweden}
\affiliation{Department of Physics, University of Connecticut, Storrs, Connecticut 06269, USA}

\author{Pedram Roushan}
\affiliation{Google AI Quantum, Mountain View, California, USA}

\author{Joris Schaltegger}
\affiliation{Nordita, Stockholm University and KTH Royal Institute of Technology, Hannes Alfv\'ens v\"ag 12, SE-106~91 Stockholm, Sweden}

%\author{Vadim Smelyanskiy}
%\affiliation{Google AI Quantum}

\author{Patrick J. Wong}
\affiliation{Nordita, Stockholm University and KTH Royal Institute of Technology, Hannes Alfv\'ens v\"ag 12, SE-106~91 Stockholm, Sweden}
\affiliation{Department of Physics, University of Connecticut, Storrs, Connecticut 06269, USA}

\date{\today}

\begin{abstract}
We investigate the interaction of a transmon qubit  with a classical gravitational field.
Exploiting the generic phenomena of the gravitational redshift and Aharonov-Bohm phase,
we show that entangled quantum states dephase with a universal rate.
The gravitational phase shift is expressed in terms of a quantum computing noise channel.
We give a measurement protocol based on a modified phase estimation algorithm which is linear in the phase drift, which is optimal for measuring the small phase that is acquired from the gravitation channel.
Additionally, we propose qubit-based platforms as quantum sensors for precision gravitometers and mechanical strain gauges as an example of this phenomenon's utility. We estimate a sensitivity for measuring the local gravitational acceleration to be $\delta g/g \sim 10^{-7}$. This paper demonstrates that classical gravitation has a non-trivial influence on quantum computing hardware, and provides an illustration of how quantum computing hardware may be utilized for purposes other than computation. While we focus on superconducting qubits, we point the universal nature of gravitational phase effects for all quantum platforms.
\end{abstract}

\maketitle

\section{Introduction}

Gravitation, understood as general relativity, and quantum theory are two pillars of modern physics. While a full quantum theory of gravitation remains elusive~\cite{kiefer,giulini2023}, 
the effects of classical gravitation on quantum systems have been well established. In 1916 Einstein proposed three classical tests of his theory of gravity: (i) the anomalous perihelion precession, seen in Mercury's orbit, (ii)  the gravitational deflection of light by the Sun, observed  by Eddington in 1919, and (iii) the gravitational redshift of the photon \cite{einstein}. All three effects have been seen experimentally now and are well established~\cite{mtw}.
%\nolinkurl{https://en.wikipedia.org/wiki/Gravitational_redshift}.  

We would like to explore the effects  of gravity on quantum information systems such as qubits. The gravitational redshift, last of three Einstein's tests, will be the topic of this discussion. 

In general the coupling between quantum systems and classical gravitation is described by quantum field theory in curved spacetime~\cite{giulini2023}. This formalism is useful for situations which ignore backreaction of matter on the metric, where the energy density of the quantum matter is small compared to the spacetime curvature.
The regime explored here is that of gravitation in the Newtonian weak-field limit to lowest order in $c^{-2}$. The dynamical degrees of freedom are considered in the low-energy regime such that special relativistic effects can be ignored and non-relativistic quantum mechanics is valid.

Quantum effects resulting from classical gravitation on neutrons have been known since the 1970s, where the Colella, Overhauser, and Werner (COW) experiment \cite{cow,RevModPhys.51.43,PhysRevA.21.1419,kiefer} has shown that interferometry of neutral particles exhibits a phase depending on the difference of the gravitational potential traversed by two paths. 
While gravitational influences on quantum systems comprised of neutrons or atoms have been well studied~\cite{cow,RevModPhys.51.43,PhysRevA.21.1419,PhysRevLett.51.1401,kiefer,abele2012,kasevich2022,overstreet2023,asenbaum2017,bothwell2022,xu2019,mueller2010}, the literature studying effects on solid state platforms is comparatively sparse, but has been growing in recent years~\cite{kiefer2005,lanzagorta},
with some interesting examples being the effect of gravitation on superconductors and quantum Hall systems~\cite{kiefer2005,hehl2004,hammad2020}.

There are also relevant practical reasons to explore the dephasing effects of gravity. The present quest to construct a practical quantum computer is one fraught with challenges~\cite{brooks2023}. A significant difficulty to manage is maintaining the phase coherence of the quantum computer's qubits, which can dephase due to noise~\cite{huang2024} whose source could be a variety of reasons. For superconducting qubits examples of noise include capacitor charge fluctuations, thermal photons, and excess quasiparticles~\cite{krantz2019}. A source which is often overlooked is that of gravitation, an effect that is generically always present. While there are methods for mitigating conventional noise channels, noise of gravitational origin cannot so straightforwardly be addressed.

In this paper we study the effects of gravitation redshift in which photon frequencies are augmented in the presence of a gravitational field~\cite{poundrebka,hafelekeating,zych2011,pikovski2015} on quantum computing hardware.
This induces a phase shift which serves to depolarize qubit states and can be interpreted as a universal noise channel. 
We examine this phenomenon and discuss the practicalities of its measurement and its effects on quantum computers.

This work contributes to a growing literature on interferometric measurements of gravity, where again existing results are based on  measurements on neutral particles~\cite{bongs2019,bothwell2022,stray2022,afanasiev2024}. 
%With the great push to construct a large scale, potentially general purpose, quantum computer, quantum sensors will emerge as a viable technology well-beforehand, and to a certain extent, already have~\cite{bongs2023}.
A single qubit with characteristic  frequency $\omega$ will acquire relative  phase shifts on the order of $\phi = \frac{g \omega \Delta x T}{c^2} $,  Eq.~\eqref{eq:abphase}, for the measurement time $T$, $g \approx \SI{9.8}{\meter\per\second\squared}$ being the Earth ($\oplus$) gravitational acceleration rate and $\Delta x$ being the typical qubit spacing. We also point out that for the typical design of qubits with natural spacing between on the order of the wavelength $x \sim \lambda$ such that the dephasing rate for vertically aligned qubits is $\dot{\phi} \sim \frac{g}{c}$. 
We also analyze ways in which qubits subject to gravity can be employed in a controlled manner as quantum sensors of gravity~\cite{bongs2023}.

We focus on superconducting qubits (transmons) as a specific platform. However, the effects we find are broadly applicable to other quantum information platforms which make use of internal energy levels to define the information basis, such as cold atoms, and hence the  use of the word {\em universal} in the title.

%Di\'{o}si-Penrose gravitational decoherence where gravity is the generic driving mechanism behind the decoherence of quantum states and the emergence of macroscopic classicality~\cite{penrose}. While thematically similar, we stress that our approach in this paper is not based on collapse models.

The structure of this paper is as follows. 
In Sec.~\ref{sec:grav_effects} we derive the standard results of the gravitational redshift and gravitational phase shift. 
In Sec.~\ref{sec:qc_implications} we discuss the implications of these effects on superconducting qubits and express the phase shift as a quantum computing noise channel.  
In Sec.~\ref{sec:measurement_protocol} we give an algorithm which is optimal for measuring the small phase induced by gravitation.
In Sec.~\ref{sec:sensors} we describe possible utility of this measurement for quantum sensing.

\section{Gravitational Effects on Quantum Systems\label{sec:grav_effects}}

The Newtonian gravitational potential is given by $\Phi(x) = - \frac{G M_\oplus}{|x_\oplus+x|}$, where $x_\oplus$ denotes the radius of Earth and $x$ the radial displacement from the surface. The local gravitational acceleration at the surface is $g \vcentcolon= \frac{G M_\oplus}{x_\oplus^2} \approx \SI{10}{\meter\per\second\squared} \equiv \SI{E3}{Gal}$.
The metric for a spherically symmetric Newtonian gravitational field in Cartesian coordinates to lowest order in $c^{-2}$ is
\begin{equation}
    \boldsymbol{g}(x)
    =
    \pmqty{\dmat{ -1 - \frac{2 \Phi(x)}{c^2}
    , 1 %- \frac{2 \Phi(x)}{c^2}
    , 1 %- \frac{2 \Phi(x)}{c^2}
    , 1 %- \frac{2 \Phi(x)}{c^2}
    }}
\label{eq:newtonianmetric}
\end{equation}
where $x$ represents the vertical distance away from the source mass. We ignore rotational effects, such as those leading to Coriolis forces, as the gravitomagnetic contribution is of order $c^{-1}$ compared to the (``gravitoelectric'') Newtonian term. In principle these terms are not entirely irrelevant as they can potentially contribute to Sagnac effects~\cite{cow,werner1979}.

To see how a non trivial space-time metric affects nonrelativistic quantum mechanics, consider that the Schr\"{o}dinger equation for a relativistic observer can be expressed as
\begin{align}
    -\frac{\hbar}{i} \pdv{\tau} \Psi &= \hat{H}_{\text{rest}} \Psi
    \\
    -\frac{\hbar}{i} u^\mu \pdv{x^\mu} \Psi &= \hat{H}_{\text{rest}} \Psi
\end{align}
where $\hat{H}_{\text{rest}}$ is the generator of proper time translations and $u^\mu = \pdv{x^\mu}{\tau}$. For a static observer, $u^\mu$ is a timelike Killing vector with only a zero component, $u^0$. From the invariant magnitude of the 4-velocity we have
\begin{align}
    g_{\mu\nu} u^\mu u^\nu &= - c^2
    \\
    \frac{u^0}{c} &= \frac{1}{\sqrt{-g_{00}}} .
\end{align}
Inserting this into the Schr\"{o}dinger equation yields
\begin{align}
%    -\frac{\hbar}{i} u^\mu \pdv{x^\mu} \Psi &= \hat{H}_{\text{rest}} \Psi
%    \\
    -\frac{\hbar}{i} u^0 \frac{\partial}{c\partial t} \Psi &= \hat{H}_{\text{rest}} \Psi
    \\
    -\frac{\hbar}{i} \pdv{t} \Psi &= \sqrt{-g_{00}} \hat{H}_{\text{rest}} \Psi
\end{align}
where now $\sqrt{-g_{00}} \hat{H}_{\text{rest}}$ is the generator of coordinate time translations.
From the metric Eq.~\eqref{eq:newtonianmetric}, we see the Hamiltonian becomes shifted by the amount $\sqrt{1+\frac{2\Phi}{c^2}} \approx 1 + \frac{\Phi}{c^2}$. For a rest Hamiltonian whose eigenenergies are given by $\hbar \omega_n$, the energy levels are shifted to $\hbar \qty(1 + \frac{\Phi}{c^2}) \omega_n$. In other words, there is a shift in the spacing of the energy levels. This is the well-known gravitational redshift~\cite{poundrebka,hafelekeating,zych2011,pikovski2015}. The gravitational redshift has measured in atomic clocks over distances of $\SI{1}{\milli\meter}$ \cite{bothwell2022,zheng2022}, where the measured frequency shift is $\mathcal{O}(10^{-19})$. This distance is $\sim\mathcal{O}(10^{-1})$ smaller than current quantum computing hardware, which implies that this effect conceivably already has relevance to current technology. The key physical phenomena associated with the effect we will discuss in the gravitational redshift, in which photon frequencies are augmented in the presence of a gravitational field~\cite{poundrebka,hafelekeating,zych2011,pikovski2015}.
More in-depth discussions of this effect in quantum systems can be found in Refs.~\cite{pikovski2015,pikovski2017}.

%This fundamental time dilation due to gravitation leads to a universal dephasing rate for quantum systems, and in particular, the hardware elements of emerging technologies such as quantum computers.

Similarly along these lines, more recently a gravitational Aharonov-Bohm effect has also been demonstrated experimentally~\cite{kasevich2022}. This effect is analogous to the conventional Aharonov-Bohm effect in electromagnetism~\cite{PhysRev.115.485} where a closed loop around a gauge potential results in a quantum phase. The phase is given by $\phi = \Delta S / \hbar$ where $\Delta S$ is the difference in the classical action over two differing trajectories between fixed start and end points. The gravitational Aharonov-Bohm phase shift is given by
\begin{align}
    \phi
    &= \frac{m}{\hbar} \int_0^T\dd t \qty[ \Phi(t,x_1) - \Phi(t,x_2) ]
    \\
    &= \frac{\omega}{c^2} \int_0^T\dd t \qty[ \Phi(t,x_1) - \Phi(t,x_2) ]
    \label{eq:gravitational_AB}
\end{align}
where $\Phi(x)$ is the Newtonian gravitational potential and $x_{1/2}(t)$ is a trajectory over a path 1 or 2 with transit time for both paths being $T$. We treat the qubits as oscillators with frequency $\omega$, and obtain an effective qubit mass from the relation $mc^2 = E = \hbar \omega$. For Mach-Zehnder parallelogram trajectories with a height difference of $\Delta x = |x_1 - x_2|$, the gravitational Aharonov-Bohm phase is
\begin{equation}
    \phi = \frac{g \omega \Delta x T}{c^2}
\label{eq:abphase}
\end{equation}
where $g$ is the local gravitational acceleration.
For sample parameters applicable to superconducting qubit hardware, we have for a $\SI{1}{\centi\meter}$ difference in path height and a transmon frequency of $\SI{10}{\giga\hertz}$ a rate of change in the phase Eq.~\eqref{eq:abphase} of $\partial_t{\phi} = \SI{E-7}{\radian/\second}$.
This quantity is naturally dependent upon the vertical extent of the interferometer, but it is also dependent upon the energy splitting of the states. Recent work~\cite{kraemer2023,tiedau2024} has shown that 
the $\SI{8.4}{\electronvolt}$ nuclear isomer state in Th-229 can be resonantly excited in Th-doped $\text{CaF}_2$ crystals with a frequency of $\sim\SI{2000}{\tera\hertz}$, which is large enough that it may be measurable with present technology. 
The drift in the phase for a coherent state of these energy levels would be significantly greater than that for transmon qubits due to the much higher frequency. Over a distance of $\SI{1}{\milli\meter}$, the rate of phase change would be $\partial_t \phi_{\text{Th-}299} \approx \SI{E-3}{\radian/\second}$.

In practice, transmon qubits are connected to each other via waveguides with a length determined by their wavelength $\lambda = c/\omega$. The spatial (vertical) distance between qubits, the $\Delta x$ which appears in Eq.~\eqref{eq:abphase}, is proportional to this wavelength,
$\Delta x = \mathcal{N} \frac{c}{\omega}$, with an engineering factor $\mathcal{N}$ that takes into account the physical characteristics of the waveguide, such as its lithographic geometry, index of refraction, etc.
This leads to the dephasing rate being expressed as
\begin{equation}
    \partial_t \phi = \mathcal{N} \frac{g}{c} .
\end{equation}
We observe the curiosity that this ratio, $\frac{g}{c}$, appears also in the Unruh temperature for a uniformly accelerated observer with acceleration $a$:
$T_U = \frac{\hbar a}{2\pi k_B c}$~\cite{unruh1976,alsing2004}. This is suggestive that a relationship between gravitationally induced dephasing and a thermal background exists. The implications of this observation are yet to be worked out.

\section{Implications for Quantum Computers\label{sec:qc_implications}}

\subsection{Effect on a single qubit} \label{sec:single_qubit_effects}
We now apply the above results to superconducting quantum computers, beginning with the effect on a single qubit. Since the rest mass of the qubit is constant, the gravitational field only couples to the internal energy, \textit{i.e.} the qubit is an ideal quantum clock. Up to a constant, we can express the internal Hamiltonian of the single qubit as $\hat{H}_0 = \hbar \omega \ketbra{1}{1}$. Neglecting the spatial extension of the qubit,\footnote{We can neglect the spatial extent of the qubit as the size of a transmon is on the order of $\SI{100}{\micro\meter}$, however the primary effects we will consider below take place on the order of centimeters.} the full Hamiltonian in the gravitational field is then simply $\hat{H} = \hbar \omega \qty(1+\frac{\Phi(x)}{c^2}) \ketbra{1}{1}$.  The gravitational coupling results in a small shift of the qubit frequency, which is automatically accounted for during calibration of the qubit frequency. In practice, the qubit is coupled to a readout resonator and a waveguide allowing implementation of single-qubit gates. However, as long as we neglect the extension of the qubit, the relative gravitational redshift is identical for the qubit and photon frequencies. Consequently, the effect of gravity is fully mitigated by calibrating the single-qubit gates. This result is unsurprising given the fact that we have assumed a qubit at rest without any spatial extension. Clearly the curvature of space-time is irrelevant in this approximation.

Changing the local gravitational field without re-calibrating the qubit frequency will result in a frequency shift of the qubit induced by the change of the gravitational potential. In particular, assume that the local potential $\Phi_0$ changes according to $\Phi_0 \rightarrow \Phi_0 + \delta \Phi$; then the qubit frequency $\omega^{(g)}$ (including the correction for $\Phi_0$) will experience a frequency shift $\omega^{(g)} \rightarrow \omega^{(g)} + \omega^{(0)} \frac{\delta \Phi}{c^2} \approx \omega^{(g)} \qty(1+\frac{\delta \Phi}{c^2})$. Here $\omega^{(0)}$ describes the ``bare'' qubit frequency in the absence of a gravitational potential.

\begin{figure}
    \centering
    \includegraphics[width=0.8\linewidth]{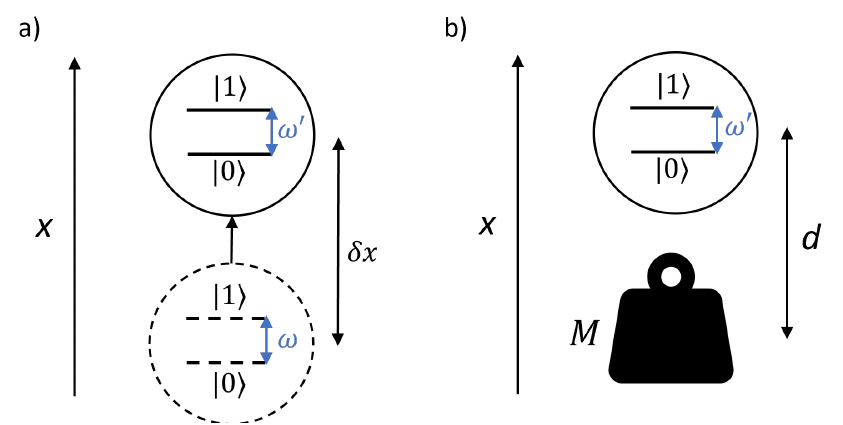}
    \caption{Depiction of two conceptual scenarios to induce a qubit frequency shift by a change in the gravitational potential. (a) Vertical movement of a qubit. (b) Massive object in the proximity of the qubit.}
    \label{fig:single_qubit_pictograms}
\end{figure}

We now consider two scenarios of how the local gravitational potential can be changed. First, we consider a vertical movement by $\delta x$ and, second, we consider the presence of an additional massive object placed in close proximity to the qubit, as shown in Fig.~\ref{fig:single_qubit_pictograms}.

%\paragraph{Vertical movement}
A vertical movement of the qubit by $\delta x$ changes the gravitational potential according to $\Phi(x_0+\delta x) \approx \Phi(x_0) \qty(1-\frac{\delta x}{x_0})$, which results in the frequency shift $\frac{\delta \omega^{(g)}}{\omega^{(g)}} \approx - \frac{\delta x}{x_0} \frac{\Phi(x_0)}{c^2} =  \frac{G M_\oplus}{x_0^2} \frac{\delta x}{c^2} \equiv \frac{g \delta x}{c^2}$, where $g$ is Earth's gravitational acceleration at distance $x_0$. Naturally, the frequency shift is tiny: for a vertical movement of $\SI{1}{\centi\meter}$ we obtain $\frac{\delta \omega^{(g)}}{\omega^{(g)}} \approx 10^{-18}$, which is far beyond experimental detection thresholds. Detecting the frequency shift due to the change in the gravitational potential would require a Ramsay-type measurement to be carried out over approximately $10^{18}$ periods or roughly one year for a qubit frequency of $\SI{10}{\giga\hertz}$. We will see, however, that a large number of coherent qubits can lead to a much better sensitivity.

%\paragraph{Massive object}
A change in the gravitational potential can also be engineered by placing the system near a massive object.
The distortion of space-time due to a massive object in the vicinity of a quantum system has been experimentally observed in matter-wave interference experiments \cite{kasevich2022}. We can estimate the effect of a massive object with mass $M$ when it is brought in close proximity to the qubit with distance $d$. The mass $M$ is responsible for an additional term in the gravitational potential, $\Phi = \Phi_0 - \frac{GM}{d} = \Phi_0 \qty(1+\frac{M}{M_\oplus} \frac{x_0}{d})$. For reasonable estimates of $M = \SI{E3}{\kilogram}$ and $d = \SI{10}{\centi\meter}$, we obtain the qubit frequency shift $\frac{\delta \omega^{(g)}}{\omega^{(g)}} \approx - \frac{1}{c^2} \frac{GM}{d} \sim - 10^{-23}$, \textit{i.e.} the effect is five orders of magnitude smaller compared to a vertical displacement. We will therefore focus on proposals involving vertical displacements in the remainder of the manuscript.

%\subsection{Two-qubit unitary gates}
\subsection{Multiple qubits}
For quantum chips with multiple qubits, in general each qubit experiences a site-dependent gravitational redshift. As in the single qubit case, the redshifts can be calibrated away on each qubit if the gravitational potential is static. In contrast, the gravitational field results in a direction-dependent effect on two-qubit unitaries, which cannot be calibrated away completely, because the gravitational field breaks the inversion symmetry. The effect of gravity on two-qubit unitaries will be explored in a separate publication; in this manuscript we will focus on the gravitational redshifts caused by a change in the local gravitational potential.
%The second order leads to direction-dependent corrections; for example, implementing an $\sqrt{i\mathrm{SWAP}}$ gate yields slightly different swap amplitudes for the upwards and downwards directions.
% Work out precise mechanism in i\mathrm{SWAP} implementation. Check whether there are terms after renormalizing all frequencies that are beyond calibration capabilities, and if so, estimate magnitudes.

In the general case, we assume that, after the calibration of the qubit frequencies, the gravitational field changes according to $\Phi_0 \rightarrow \Phi_0 + \delta \Phi_k$, where the change of the gravitational potential can be site dependent. For example, if the quantum computer is rotated, each qubit experiences a change in the gravitational potential depending on its vertical displacement, resulting in a frequency shift according to $\delta \omega_k = \frac{\delta \Phi_k \omega_k}{c^2}$. A computational basis state $\qty{\ket{j}}$, $j\in \qty{0,1}^n$ will now experience a gravitational phase shift based on the frequency shifts of all excited qubits:
\begin{equation}
    \label{eq:grav_phase_shift}
    \phi_{\ket{j}} = -\frac{t}{c^2}\sum_{k=1}^{n} \expval{P_k}{j} \omega_k \delta \Phi_k,
\end{equation}
with the operators $P_k = (\mathbbm{1}_2)^{\otimes^{k-1}} \otimes 
\pmqty{
    0 & 0 \\
    0 & 1
}
\otimes (\mathbbm{1}_2)^{\otimes^{n-k}}$ which project states onto $\ket{1}$ at site $k$. This is the qubit representation of the Aharonov-Bohm phase Eq.~\eqref{eq:gravitational_AB}.
We can therefore express the gravitational phase shift as a dephasing channel in terms of Kraus operators $\Sigma$, 
\begin{align}
\Lambda : \rho &\mapsto \Sigma(t)\rho \Sigma^\dagger(t), \\
\Sigma(t) &= \bigotimes_{k=1}^n \Sigma_k(t), \\
\Sigma_k(t) &= 
\pmqty{
1 & 0 
\\
0 & e^{i \theta_k}
},
\end{align}
where the dephasing angle for each qubit is $\theta_k \equiv \frac{t}{c^2} \delta \Phi_k \omega_k$. This representation of the gravitational phase shift facilitates the analysis of the effects of this phase shift on quantum circuits.
Since this is a pure dephasing effect, in principle it can be compensated for using known techniques~\cite{krantz2019,bluvstein2024}. However, our present goal is to actually read out this information, not erase it.
What this means is that, at the algorithmic level, any error correction code should be designed to correct for other forms of error, but not the effect from gravity. Specifics of an algorithm designed to achieve this depends on details of qubit operation and is beyond the scope of the present manuscript.
%

%\subsubsection{Rotated quantum computer}
As an illustration, we consider a specific example, where a quantum computer is calibrated in a horizontal alignment and subsequently rotated to a vertical alignment around its center of gravity. In this case, the change in the local gravitational potential for each qubit is determined by the vertical displacement $x_k$ from the rotation, which results in the frequency shifts $\omega_k \to \qty( 1 - \frac{g x_k}{c^2} ) \omega_k$. In this case, the dephasing angles are $\theta_k = - \frac{gt}{c^2} \omega_k x_k$.

%twirling\cite{horodecki1999,nielsen2002}

\section{Measuring the gravitational phase shift} \label{sec:measurement_protocol}
The gravitational phase shift can be maximally demonstrated using an interferometry protocol with a Greenberger-Horne-Zeilinger (GHZ) state~\cite{greenberger1989}. In our case, the optimal superposition state is not always a GHZ state but is closely related. In particular, we split the qubits into two groups according to the sign of their gravitational phase shift. The maximal signature of the gravitational phase shift is then achieved by a superposition of the state with all qubits with positive phase shift excited, and the state with all qubits with negative phase shift excited. We introduce the notation 
\begin{align}
    \ket{+} &\vcentcolon= \bigotimes_{\{j|\theta_j \geq 0\}} \sigma_j^+ \ket{\vec{0}} \\
    \ket{-} &\vcentcolon= \bigotimes_{\{j|\theta_j < 0\}} \sigma_j^+ \ket{\vec{0}}
\end{align}
to denote the states which have all qubits with positive (negative) gravitational phase shifts excited. The desired state is then the superposition $\ket{\Psi} = \frac{1}{\sqrt2} \left( \ket{+} + \ket{-} \right)$. Schematically this can be depicted as
\begin{equation}
\label{eq:GHZ_state_schematic}
    \ket{\Psi} = \frac{1}{\sqrt2}
    \ket{\begin{matrix}
        \bullet \\[-0.27cm] \vdots \\[-0.2cm] \bullet \\[-0.2cm] \circ \\[-0.27cm] \vdots \\[-0.2cm] \circ
    \end{matrix}}
    +
    \frac{1}{\sqrt2}
    \ket{\begin{matrix}
        \circ \\[-0.27cm] \vdots \\[-0.2cm] \circ \\[-0.2cm] \bullet \\[-0.27cm] \vdots \\[-0.2cm] \bullet
    \end{matrix}}
\end{equation}
where $\bullet$ represents $\ket{1}_j$ and $\circ$ represents $\ket{0}_j$ in the computational basis.

For example, if the phase shifts are achieved by rotating the quantum computer from a horizontal to a vertical alignment, the sign of the phase shift depends on whether the vertical displacement $x_k$ is positive or negative, that is, qubits in the upper half of the chip will experience a negative phase shift while qubits in the lower half experience a positive phase shift. On the other hand, if the gravitational phase shifts come from a local change in the gravitational acceleration $g\rightarrow g+ \delta g$, all qubits exhibit a gravitational phase with the same sign and the desired state is indeed a GHZ state.

After creating the optimal superposition state on the quantum computer, the gravitational phase accumulates over time according to Eq.~\eqref{eq:grav_phase_shift}, which we label as $\phi_\pm$ respectively. Schematically, the phase accumulation is expressed as
\begin{equation}
    \ket{\Psi(t)} = 
    \frac{1}{\sqrt2}
    \ket{\begin{matrix}
        \bullet \\[-0.27cm] \vdots \\[-0.2cm] \bullet \\[-0.2cm] \circ \\[-0.27cm] \vdots \\[-0.2cm] \circ
    \end{matrix}}
    e^{i \phi_{+}(t)}
    +
    \frac{1}{\sqrt2}
    \ket{\begin{matrix}
        \circ \\[-0.27cm] \vdots \\[-0.2cm] \circ \\[-0.2cm] \bullet \\[-0.27cm] \vdots \\[-0.2cm] \bullet
    \end{matrix}}
    e^{i \phi_{-}(t)} .
\end{equation}
Clearly, by splitting the qubits according to the sign of their phase shift, the difference $\Delta \phi = \phi_+ - \phi_-$ is maximized, which will serve as a signature for the gravitational phase shift.

\subsection{Measurement protocol}
Here we propose a simple measurement protocol based on Ref.~\cite{mohammadbagherpoor2019} that produces an experimental signature linear in $\Delta \phi$. The measurement protocol is a variation of the standard phase estimation algorithm, which is normally used to measure an eigenvalue of a unitary \cite{Kitaev2002}. The standard algorithm results in a measurement outcome that is proportional to $\cos(\Delta\phi)$, which is quadratic in $\Delta\phi$ for small angles. As the phase we aim to measure is already quite small, a measurement $\propto \cos(\Delta\phi) \sim \Delta\phi^2$ would be suboptimal. We therefore seek a more optimal algorithm such that the measurement outcome is linear in $\Delta\phi$.
We adjust the standard algorithm to measure the accumulated gravitational phase shift of the GHZ-like superposition state introduced in the previous section, and to observe a signature scaling linearly in the phase instead of the usual cosine signature, which of course exhibits much better sensitivity for small phases. The measurement protocol consists of the following steps: First, we initialize a state by entangling the GHZ-like state with an ancilla qubit in a suitable way. Second, we wait for a certain time $t$, during which the state experiences gravitational dephasing. Finally, the ancilla qubit is rotated and measured in the computational basis.

\subsubsection{Quantum algorithm}
In the first step, we create the entangled state
$\frac{1}{\sqrt2}\left[ \ket{0} \otimes \ket{-} + i \ket{1} \otimes \ket{+}\right]$ starting from the initial state $\ket{0} \otimes \ket{\vec{0}}$, where we explicitly isolate the ancilla qubit in our notation. This is easily achieved by the following steps:
\begin{enumerate}
    \item $H \otimes U_-$ \newline
    Here, $H$ is the Hadamard gate and $U_-$ denotes the generator of the $\ket{-}$ state, i.e. $U_- = \bigotimes_{\{j|\theta_j < 0\}} \sigma_j^+$.
    \begin{equation}
        \longrightarrow \frac{1}{\sqrt2} \left( \ket{0} + \ket{1} \right) \otimes \ket{-}
    \end{equation}

    \item $S \otimes \mathbb 1$ \newline
    The $S$-gate is given by $S = \ketbra{0}{0} + i \ketbra{1}{1}$, applying it to the ancilla qubit is required for a linear signature in the end.
    \begin{equation}
        \longrightarrow \frac{1}{\sqrt2} \left( \ket{0} + i \ket{1} \right) \otimes \ket{-}
    \end{equation}

    \item $C-(U_-^\dagger U_+)$ \newline
    Next we apply the controlled unitary $U_-^\dagger U_+$ that destroys the $\ket{-}$ state and generates the $\ket{+}$ state instead. This generates the desired entangled state:
    \begin{equation}
        \longrightarrow \frac{1}{\sqrt2} \left( \ket{0} \otimes \ket{-} + i \ket{1} \otimes \ket{+} \right)
    \end{equation}

    \item Wait for time $t$ \newline
    Next, we let the gravitational phase accumulate for some time $t$ (we ignore the gravitational phase shifts on the ancilla qubit here).
    \begin{equation}
        \longrightarrow \frac{1}{\sqrt2} \left( \ket{0} \otimes \ket{-} e^{i \phi_-(t)} + i \ket{1} \otimes \ket{+} e^{i \phi_+(t)} \right)
    \end{equation}

    \item $H \otimes \mathbb{1}$ \newline
    The final step consists of applying a Hadamard gate and subsequent measurement of the ancilla qubit, where we use $\Delta \phi = \phi_+ - \phi_-$.
    \begin{equation}
    \begin{aligned}
        \longrightarrow \frac{e^{i \phi_-(t)}}{2} \big[ 
        &\ket{0} \otimes \left( \ket{-} + i e^{i \Delta\phi} \ket{+}\right) \\
        + &\ket{1} \otimes \left( \ket{-} - i e^{i \Delta\phi} \ket{+}\right)
        \big]
    \end{aligned}
    \end{equation}
    
    \end{enumerate}

\begin{figure}
    \centering
    \begin{comment}
    \begin{quantikz} \label{fig:circ}
    	\lstick{$\ket{0}$} & \gate{H} & \gate{S} & \ctrl{1} & \ \ldots \ \gategroup[3,steps=1,style={dashed,rounded corners},label style={label position=below,anchor=north,yshift=-0.2cm}]{\textit{t}} & \gate{H} & \meter{}
        \\
    	\lstick[2]{$\ket{\vec{0}}$} & \ghost{U_-} & & \gate[2]{U_-^\dagger U_+} & \ \ldots \ & & 
        \\
    	& \gate{U_-} & & & \ \ldots \ & &
    \end{quantikz}
    \end{comment}
    \includegraphics[]{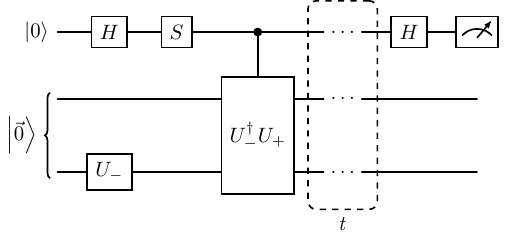}
    \caption{Circuit diagram of the gravitational phase measurement protocol, where the uppermost line denotes the ancilla qubit and the other qubits have been split according to the sign of their gravitational phase.}
    \label{fig:circuit_phase_measurement}
\end{figure}

Measuring the ancilla qubit in the computational basis then results in the following probabilities:
\begin{align}
    P(0) &= \norm{\frac12 (1 + i e^{i \Delta\phi})}^2 = \frac12 - \frac12 \sin(\Delta\phi), \\
    P(1) &= \norm{\frac12 (1 - i e^{i \Delta\phi})}^2 = \frac12 + \frac12 \sin(\Delta\phi).
\end{align}
As desired, we obtain a linear signature in the phase difference $\Delta \phi$, which has a much better sensitivity for small angles than the result of the typical phase estimation algorithm~\cite{kitaev1995,Kitaev2002}, which gives probabilities scaling with $\cos(\Delta\phi)$. The circuit diagram of the phase measurement protocol is depicted in Fig.~\ref{fig:circuit_phase_measurement}.

The phase estimation algorithm can be parallelized efficiently by using additional ancilla qubits and applying the inverse quantum Fourier transform \cite{Kitaev2002,mohammadbagherpoor2019}. We believe that the algorithm proposed here could be adjusted analogously by measuring the ancilla qubits in regular intervals. However, the usefulness of such a procedure is reduced because the coherence time $T_{c}$ of the quantum computer limits the phase accumulation time between ancilla measurements to $t \sim \frac{T_c}{m}$ for $m$ ancillas, whereas the procedure with only a single ancilla qubit can make use of the full $T_c$ to maximize the accumulated phase shift.

\subsection{Proof-of-concept measurement of the gravitational phase shift}
To our knowledge, the gravitational redshift has never been experimentally observed for massless quantum excitations. The most prominent experimental candidates for such measurements are based on photon interference, but it is in practice very hard to accumulate a sufficiently large gravitational phase shift. With qubits, the possibility of controlling many degrees of freedom with high precision for a comparatively long time opens up novel scaling opportunities compared to photon interference experiments. We are therefore convinced that quantum computers will play an important part in experimental tests of quantum gravity in the future. The experimental evidence of the gravitational phase shift on a quantum computer will constitute the first step towards this goal. Here we explore the possibility and requirements of such an experiment.

In most applications, due to the additive property of the gravitational phase, we expect that the total accumulated phase scales linearly in the number of qubits $n$, for example, from a change in the gravitational acceleration. However, using a more elaborate procedure, it is possible to achieve a quadratic scaling with the number of qubits, which facilitates the experimental confirmation of the gravitational phase shift on near-term quantum computers. In particular, we consider a 1D strip of qubits, which is calibrated in a horizontal alignment and subsequently rotated around its center of gravity to a vertical alignment. The individual qubits then experience a frequency shift due to the vertical displacement $x_k$ according to $\omega_k \to \qty( 1 - \frac{g x_k}{c^2} ) \omega_k$. We then find the maximally sensitive superposition as depicted in Eq.~\eqref{eq:GHZ_state_schematic}, that is, the $\ket{+}$ is generated by exciting all qubits in the lower half of the chip, while the $\ket{-}$ state is generated by exciting all qubits in the upper half of the chip. The cumulative phase difference is then given by
\begin{equation}
    \Delta \phi (t) = \frac{gt}{c^2} \Big[ \sum_{ \{ k | x_k >0 \} } \omega_k x_k - \sum_{ \{ k | x_k < 0 \} } \omega_k x_k
    \Big].
\label{eq:cumulative_phase_difference}
\end{equation}
Assuming an even number of homogeneously spaced qubits, the vertical coordinates after the rotation are given by $x_k = \frac{n + 1 - 2k}{2} \ell$, where $\ell$ denotes the distance between neighboring qubits. The total phase difference is then given by
\begin{equation}
    \Delta \phi (t) \approx \frac{g\bar{\omega}\ell n^2 t}{4c^2},
\end{equation}
where $\bar{\omega}$ is the average qubit frequency and we have neglected the qubit detuning, as it is a second-order effect. In this procedure, the rotation by $\frac{\pi}{2}$ creates an average vertical displacement scaling with the linear dimension $L = n \ell$ of the chip, which contributes an additional factor of $n$ in addition to the cumulative property of the gravitational phase in the GHZ-like state. For a 1D-architecture of the chip, we therefore obtain a total scaling of $\Delta \phi \propto n^2$; an equivalent procedure on a 2D chip results in a scaling of $\propto n^{3/2}$, because the linear dimension of the chip is $L = \sqrt{n} \ell$ in this case.

Assuming a resolution of $\SI{0.1}{\radian}$ for the phase measurement protocol, we can estimate the required properties of the chip to measure the gravitational phase shift. Current superconducting quantum computers have interqubit distances of approximately $\ell \sim \SI{1}{\milli\meter}$ and frequencies of $\omega \sim \SI{10}{\giga\hertz}$. We do not expect those numbers to change drastically, since the length of the waveguides connecting the qubits is fundamentally limited by the wavelength of the microwave photons. We therefore do not expect the order of magnitude of $\bar{\omega} \ell$ to change drastically with future generations of quantum computers. Maximal coherence times of superconducting qubits currently reach close to $T_c \sim \SI{E-3}{\second}$. If those specifications remain unchanged, a 1D geometry would require about $n \sim 10^5$ qubits, which corresponds to $L \sim \SI{100}{\meter}$. Clearly, designing such a chip is rather unrealistic in the near future; hence it is necessary to improve on aspects other than the number of qubits as well. Most likely, coherence times will improve by several orders of magnitude using error-correction codes, and could be further increased by special-purpose chips to reduce error sources. If a coherence time of $T_c \sim \SI{1}{\second}$ is achieved, the required number of qubits is reduced to $n \sim 5000$, which corresponds to $L \sim \SI{5}{\meter}$, a more realistic scale albeit still very large. It is worth pointing out that 2D geometries have a better scaling with the linear dimension of the chip than 1D geometries. Assuming again a coherence time of $T_c \sim \SI{1}{\second}$, the gravitational phase can be measured with $n \sim 10^6$ qubits, which corresponds to a 2D chip of size $\SI{1}{\meter} \times \SI{1}{\meter}$.

\section{Gravitational Quantum Sensors\label{sec:sensors}}

The dephasing introduced by the background gravitational field may be exploited as a means of developing a device which serves as a precision sensor. For a known fixed qubit frequency and processing time, we see from the shift $\omega_k \to \qty( 1 - \frac{g x_k}{c^2} ) \omega_k$ that the remaining degrees of freedom are the local gravitational acceleration $g$ and the qubit displacement $\delta x$.
Measurements based on variations in $g$ enable the phase shift to be used to serve as the basis of a gravimeter. Measurements based on variations in the relative qubit displacements $\delta x$ could be employed to measure mechanical strain.

\subsection{Application as a gravitometer}
In this section we provide estimates for the sensitivity towards changes in the gravitational acceleration based on the measurement protocol introduced in Sec.~\ref{sec:measurement_protocol}. In particular, we consider a change $g \rightarrow g + \delta g$, which corresponds to a change in the local gravitational potential of $\delta \Phi = - x_\oplus \delta g$, where $x_\oplus$ is the radius of the Earth. Here we assume that the change of the gravitational acceleration does not affect the center of gravity of the Earth, and therefore $x_\oplus$ is unchanged. This scenario is different to the proposal discussed in Sec.~\ref{sec:single_qubit_effects}, where a large mass is placed in close proximity of the qubits. In particular, the additional mass results in a small shift in the effective center of gravity probed by the qubit, which results in a much smaller effect on the gravitational potential. A change in the gravitational acceleration results in a frequency shift for each qubit according to $\delta \omega_k = - \frac{x_\oplus \delta g \omega_k}{c^2}$. Since the frequency shift has an identical sign for each qubit, the maximally sensitive superposition state is the GHZ state. The total phase picked up during a measurement of time $t$ is given by
\begin{equation}
    \Delta \phi (t) = \frac{x_\oplus \delta g t}{c^2} \bar{\omega} n,
\end{equation}
where $\bar{\omega}$ is the average qubit frequency. 

Assuming that a phase of $\SI{0.1}{\radian}$ can be resolved, we can estimate the sensitivity towards $\delta g$. For near-term qubit architectures, we use the numbers $n \sim 10^3, \ T_c = \SI{E-3}{\second}$ and $\bar{\omega} = \SI{10}{\giga\hertz}$, which results in $\delta g \sim \SI{0.1}{\meter\per\second\squared}$ or in relative terms $\frac{\delta g}{g} \sim 10^{-2}$.

However, it is in principle fairly easy to improve the sensitivity by scaling up the number of qubits and increasing the coherence time. In particular, an optimized chip built specifically for application as a gravitometer is not required to be a universal quantum computer, and only requires the implementation of entangling gates with the ancilla qubit. All other operations used in the measurement protocol are single-qubit operations. A special-purpose chip could therefore eliminate a large proportion of decoherence events. Furthermore, error correction codes promise massively improved coherence times for logical qubits. We are therefore convinced that decoherence can reach orders of seconds in the future, which enables a much higher sensitivity. Assuming $T_c \sim \SI{1}{\second}$ and $n \sim 10^5$, the relative sensitivity towards changes in the gravitational acceleration reaches $\frac{\delta g}{g} \sim 10^{-7}$. In comparison, state-of-the-art commercial gravimeters advertise a sensitivity of $\SI{0.1}{\micro Gal} \approx 10^{-10} g$ \cite{scintrex}. 

We remark that the gravitational phase on qubits is a direct probe of the gravitational potential, rather than a probe of the gravitational acceleration as used in conventional gravimeters. This is potentially advantageous to isolate systematic errors from nearby heavy objects. For example, the scenario depicted in Fig.~\ref{fig:single_qubit_pictograms}~b) leads to a relative change $\frac{\delta g}{g} \approx 10^{-6}$ for $d = \SI{0.1}{\meter}$ and $M = \SI{E3}{\kilogram}$, but the change in the gravitational potential is suppressed by a factor of $\frac{d}{x_{\oplus}}\sim 10^{-8}$ when compared to a change in $g$ from Earth's density fluctuations (that is, without changing the center-of-mass distance $x_{\oplus}$).

\subsection{Application as a mechanical strain gauge}

\begin{figure}[ht!]
\includegraphics[]{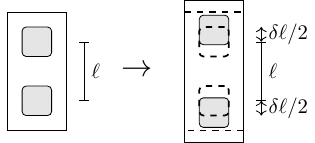}
\caption{Change in the displacement of qubits $\ell \to \ell + \delta \ell$ under an applied force.}
\label{fig:strain}
\end{figure}

In addition to the local gravitational potential, we propose that the gravitational dephasing we describe can also be employed to measure deviations in strain. 
The strain considered here will be strain along the axis which is along the gradient of the gravitational field. 
This mechanical strain in a qubit device would manifest as a shift in the accumulated gravitational phase. (See Fig.~\ref{fig:strain}.)

Once again, we take the GHZ state as the maximally sensitive superposition state.
The cumulative phase would be obtained as described in the previous section; however, the position of each qubit $x_k$ would experience a shift in displacement. For simplicity, we assume that the strain manifests as a uniform shift in the interqubit distances $\ell$ throughout the device. The position of each qubit then undergoes a shift of $x_k \sim \ell \to \ell + \delta \ell$. The effect on the phase shift of this displacement is
\begin{equation}
    \Delta\phi(t) = \frac{g \ell \bar\omega n t}{c^2} \left( 1 + \frac{\delta \ell}{\ell} \right) .
\end{equation}
The effect of strain represents a small correction to the already small phase shift acquired by the gravitational effect. 
Existing examples of precision strain gauges which are based on MEMS devices have a sensitivity of $\frac{\delta \ell}{\ell} \sim \SI{E-6}{}$~\cite{ramirez2020}. This is well below what would be measurable assuming a phase resolution of $\SI{0.1}{\radian}$. A qubit based device would therefore need to be sensitive to phase shifts of this order of magnitude in order to be comparable to existing technologies.

\section{Conclusion\label{sec:conclusion}}

In this paper we derived a universal dephasing channel for qubits of gravitational origin. We have shown that the gravitational redshift opens up a new avenue for the introduction of noise in quantum circuits. In addition to the vertical dimension of the circuit, this effect scales with the processing time as well as the number of qubits involved. 
As we emphasize with our title, this phenomenon is universal, as gravity couples equivalently to all matter.
While we have focused on superconducting solid state qubits in this work, the concepts apply equally well to other forms of quantum information platforms which are based on the computational basis being defined by an energy level splitting.
The gravitational phase we report is proportional to the vertical displacement between the qubits, the energy level splitting defining the computational basis, and the integration time of the circuit, which is constrained by the coherence time of the qubits. For platforms such as trapped ions, these values can be much larger than those of superconducting qubits, so in principle the effect could be larger than that for the superconducting qubits discussed here.

As a sample application of this phenomenon, we proposed a means of exploiting this dephasing channel in a positive sense by demonstrating how this could form the basis for gravitometers or mechanical strain sensors. While our proposals based on current qubit hardware are not competitive compared to present industry standards for gravity or strain sensors, they do illustrate a new avenue for quantum sensing. In particular, we note that our measurement protocol is linear in the gravitational phase, rather than quadratic, such as in other proposals for quantum-based sensing of gravitational acceleration~\cite{cepollaro2023}. Our proposal also extends the paradigm of quantum-based gravitational sensors beyond that of atom interferometry, which is the current standard for such devices~\cite{bongs2019}. As demonstrated above, this paradigm now includes solid-state-based devices.

With present technology, qubits require a somewhat large physical area. Qubits based on solid-state hardware have a typical spacing size of $\SI{1}{\milli\meter}$, compared to classical computer hardware which have circuit elements on nanometer spacing. A quantum computer with thousands of qubits may conceivably require qubits which are a far distance apart. In order to construct a quantum computer with a sufficient number of logical qubits for practical use, the number of physical qubits needed may exceed the number which can be stored in a single hardware device, such as a cryostat. The whole device would then require the linking of multiple refrigeration systems, which may be at different gravitational potentials. Entanglement between qubits separated by a distance of $\SI{30}{\meter}$ has already been demonstrated \cite{storz2023}, showing that long distance entanglement is technologically feasible to be implemented. Should this distance be oriented against  a gravitational potential, the effect would be considerably larger compared to the effects we discussed in this paper, which were on the order of $\SI{1}{\centi\meter}$.

In the context of how this effect manifests in quantum computing circuits, we have observed that this effect affects the fidelity of two-qubit gates, involving qubits at different heights. For example, implementing the $\sqrt{i\mathrm{SWAP}}$ gate results in a swap angle which is different for upwards and downwards directions due to the gravitational field~\cite{schaltegger2023}. We will analyze this phenomenon in greater detail in a forthcoming publication.

We only considered the case where there is a vertical displacement involved across a quantum circuit. We mentioned that other means of altering the gravitational field where a qubit sits can be achieved for example by placing a large mass near the qubit. While we did not analyze this situation as it is orders of magnitude smaller than the vertical displacement case, it is possible that such effects will need to be taken into account for space based satellite quantum devices~\cite{bassi2022}, which would experience varying a mass due to the inhomogeneities in Earth's mass distribution, yielding a time-dependent gravitational Aharonov-Bohm phase~\cite{chiao2023}.

Our preceding discussion considered the effects of gravitation on the coherence of superconducting qubits where the transmon energy levels are shifted in accordance to the gravitational redshift. A different effect would be present for qubits based on spin states of spinors, which would involve the coupling between the spin and the gravitational field. In this case, Wigner rotation due to coupling between the rotation of the Earth and the qubit's spin would be a potential source of dephasing~\cite{lanzagorta}. Another contribution may arise from gravitomagnetic effects due to a rotating mass, which manifest in the $g_{0i}$ components of the metric, playing the role of a vector like potential. While these contributions would be of order $c^{-1}$ compared to effects arising from the scalar potential, it has been proposed that Sagnac effects arising from the rotation of Earth would be measurable in quantum systems~\cite{werner1979,toros2022}. Additionally, it is possible that a spin qubit would acquire a phase due to space-time torsion, as torsion couples to spin angular momentum~\cite{RevModPhys.48.393,bakke2009}. A device capable of such a measurement would be a novel contribution to the gravitational community as there is at present no empirical measurements of space-time torsion.
%We leave analysis of spin effects for future work.

Unlike noise of electromagnetic, thermal, or mechanical origin, gravitation is an effect which cannot be shielded against. While there are proposals for architectural designs which can be used to mitigate this effect~\cite{crowder2023}, the influence of gravity will always be present and will only become more impactful as device hardware increases in size.

%\section{Acknowledgements}
\begin{acknowledgements}
We are grateful to G. Aeppli,  I. Drozdov, X. Qi,  V. Khemani, D. Schuster,  V. Smelyanskiy, and M. Zych 
for useful discussions. We thank the reviewers for their helpful comments. This work was supported by the Knut and Alice Wallenberg Foundation Grant No. KAW 2019.0068, the European Research Council under the European Union Seventh Framework Grant No. ERS-2018-SYG 810451 HERO, and the University of Connecticut.
\end{acknowledgements}

\bibliography{references}

%Control: production of eprint (1) enabled
\begin{thebibliography}{52}%
\makeatletter
\providecommand \@ifxundefined [1]{%
 \@ifx{#1\undefined}
}%
\providecommand \@ifnum [1]{%
 \ifnum #1\expandafter \@firstoftwo
 \else \expandafter \@secondoftwo
 \fi
}%
\providecommand \@ifx [1]{%
 \ifx #1\expandafter \@firstoftwo
 \else \expandafter \@secondoftwo
 \fi
}%
\providecommand \natexlab [1]{#1}%
\providecommand \enquote  [1]{``#1''}%
\providecommand \bibnamefont  [1]{#1}%
\providecommand \bibfnamefont [1]{#1}%
\providecommand \citenamefont [1]{#1}%
\providecommand \href@noop [0]{\@secondoftwo}%
\providecommand \href [0]{\begingroup \@sanitize@url \@href}%
\providecommand \@href[1]{\@@startlink{#1}\@@href}%
\providecommand \@@href[1]{\endgroup#1\@@endlink}%
\providecommand \@sanitize@url [0]{\catcode `\\12\catcode `\$12\catcode `\&12\catcode `\#12\catcode `\^12\catcode `\_12\catcode `\%12\relax}%
\providecommand \@@startlink[1]{}%
\providecommand \@@endlink[0]{}%
\providecommand \url  [0]{\begingroup\@sanitize@url \@url }%
\providecommand \@url [1]{\endgroup\@href {#1}{\urlprefix }}%
\providecommand \urlprefix  [0]{URL }%
\providecommand \Eprint [0]{\href }%
\providecommand \doibase [0]{https://doi.org/}%
\providecommand \selectlanguage [0]{\@gobble}%
\providecommand \bibinfo  [0]{\@secondoftwo}%
\providecommand \bibfield  [0]{\@secondoftwo}%
\providecommand \translation [1]{[#1]}%
\providecommand \BibitemOpen [0]{}%
\providecommand \bibitemStop [0]{}%
\providecommand \bibitemNoStop [0]{.\EOS\space}%
\providecommand \EOS [0]{\spacefactor3000\relax}%
\providecommand \BibitemShut  [1]{\csname bibitem#1\endcsname}%
\let\auto@bib@innerbib\@empty
%</preamble>
\bibitem [{\citenamefont {Kiefer}(2012)}]{kiefer}%
  \BibitemOpen
  \bibfield  {author} {\bibinfo {author} {\bibfnamefont {C.}~\bibnamefont {Kiefer}},\ }\href {https://doi.org/10.1093/acprof:oso/9780199585205.001.0001} {\emph {\bibinfo {title} {{Quantum Gravity}}}}\ (\bibinfo  {publisher} {Oxford University Press},\ \bibinfo {year} {2012})\BibitemShut {NoStop}%
\bibitem [{\citenamefont {Giulini}\ \emph {et~al.}(2023)\citenamefont {Giulini}, \citenamefont {Gro{\ss}ardt},\ and\ \citenamefont {Schwartz}}]{giulini2023}%
  \BibitemOpen
  \bibfield  {author} {\bibinfo {author} {\bibfnamefont {D.}~\bibnamefont {Giulini}}, \bibinfo {author} {\bibfnamefont {A.}~\bibnamefont {Gro{\ss}ardt}},\ and\ \bibinfo {author} {\bibfnamefont {P.~K.}\ \bibnamefont {Schwartz}},\ }\bibinfo {title} {Coupling quantum matter and gravity},\ in\ \href {https://doi.org/10.1007/978-3-031-31520-6_16} {\emph {\bibinfo {booktitle} {Modified and Quantum Gravity: From Theory to Experimental Searches on All Scales}}},\ \bibinfo {editor} {edited by\ \bibinfo {editor} {\bibfnamefont {C.}~\bibnamefont {Pfeifer}}\ and\ \bibinfo {editor} {\bibfnamefont {C.}~\bibnamefont {L{\"a}mmerzahl}}}\ (\bibinfo  {publisher} {Springer International Publishing},\ \bibinfo {address} {Cham},\ \bibinfo {year} {2023})\ pp.\ \bibinfo {pages} {491--550}\BibitemShut {NoStop}%
\bibitem [{\citenamefont {Einstein}(1916)}]{einstein}%
  \BibitemOpen
  \bibfield  {author} {\bibinfo {author} {\bibfnamefont {A.}~\bibnamefont {Einstein}},\ }\bibfield  {title} {\bibinfo {title} {Die grundlage der allgemeinen relativitätstheorie},\ }\href {https://doi.org/https://doi.org/10.1002/andp.19163540702} {\bibfield  {journal} {\bibinfo  {journal} {Ann. Phys.}\ }\textbf {\bibinfo {volume} {7}},\ \bibinfo {pages} {769} (\bibinfo {year} {1916})}\BibitemShut {NoStop}%
\bibitem [{\citenamefont {Meisner}\ \emph {et~al.}(1973)\citenamefont {Meisner}, \citenamefont {Thorne},\ and\ \citenamefont {Wheeler}}]{mtw}%
  \BibitemOpen
  \bibfield  {author} {\bibinfo {author} {\bibfnamefont {C.~W.}\ \bibnamefont {Meisner}}, \bibinfo {author} {\bibfnamefont {K.~S.}\ \bibnamefont {Thorne}},\ and\ \bibinfo {author} {\bibfnamefont {J.~A.}\ \bibnamefont {Wheeler}},\ }\href@noop {} {\emph {\bibinfo {title} {Gravitation}}}\ (\bibinfo  {publisher} {W. H. Freeman and Company},\ \bibinfo {address} {San Francisco},\ \bibinfo {year} {1973})\BibitemShut {NoStop}%
\bibitem [{\citenamefont {Colella}\ \emph {et~al.}(1975)\citenamefont {Colella}, \citenamefont {Overhauser},\ and\ \citenamefont {Werner}}]{cow}%
  \BibitemOpen
  \bibfield  {author} {\bibinfo {author} {\bibfnamefont {R.}~\bibnamefont {Colella}}, \bibinfo {author} {\bibfnamefont {A.~W.}\ \bibnamefont {Overhauser}},\ and\ \bibinfo {author} {\bibfnamefont {S.~A.}\ \bibnamefont {Werner}},\ }\bibfield  {title} {\bibinfo {title} {Observation of gravitationally induced quantum interference},\ }\href {https://doi.org/10.1103/PhysRevLett.34.1472} {\bibfield  {journal} {\bibinfo  {journal} {Phys. Rev. Lett.}\ }\textbf {\bibinfo {volume} {34}},\ \bibinfo {pages} {1472} (\bibinfo {year} {1975})}\BibitemShut {NoStop}%
\bibitem [{\citenamefont {Greenberger}\ and\ \citenamefont {Overhauser}(1979)}]{RevModPhys.51.43}%
  \BibitemOpen
  \bibfield  {author} {\bibinfo {author} {\bibfnamefont {D.~M.}\ \bibnamefont {Greenberger}}\ and\ \bibinfo {author} {\bibfnamefont {A.~W.}\ \bibnamefont {Overhauser}},\ }\bibfield  {title} {\bibinfo {title} {Coherence effects in neutron diffraction and gravity experiments},\ }\href {https://doi.org/10.1103/RevModPhys.51.43} {\bibfield  {journal} {\bibinfo  {journal} {Rev. Mod. Phys.}\ }\textbf {\bibinfo {volume} {51}},\ \bibinfo {pages} {43} (\bibinfo {year} {1979})}\BibitemShut {NoStop}%
\bibitem [{\citenamefont {Staudenmann}\ \emph {et~al.}(1980)\citenamefont {Staudenmann}, \citenamefont {Werner}, \citenamefont {Colella},\ and\ \citenamefont {Overhauser}}]{PhysRevA.21.1419}%
  \BibitemOpen
  \bibfield  {author} {\bibinfo {author} {\bibfnamefont {J.~L.}\ \bibnamefont {Staudenmann}}, \bibinfo {author} {\bibfnamefont {S.~A.}\ \bibnamefont {Werner}}, \bibinfo {author} {\bibfnamefont {R.}~\bibnamefont {Colella}},\ and\ \bibinfo {author} {\bibfnamefont {A.~W.}\ \bibnamefont {Overhauser}},\ }\bibfield  {title} {\bibinfo {title} {Gravity and inertia in quantum mechanics},\ }\href {https://doi.org/10.1103/PhysRevA.21.1419} {\bibfield  {journal} {\bibinfo  {journal} {Phys. Rev. A}\ }\textbf {\bibinfo {volume} {21}},\ \bibinfo {pages} {1419} (\bibinfo {year} {1980})}\BibitemShut {NoStop}%
\bibitem [{\citenamefont {Bonse}\ and\ \citenamefont {Wroblewski}(1983)}]{PhysRevLett.51.1401}%
  \BibitemOpen
  \bibfield  {author} {\bibinfo {author} {\bibfnamefont {U.}~\bibnamefont {Bonse}}\ and\ \bibinfo {author} {\bibfnamefont {T.}~\bibnamefont {Wroblewski}},\ }\bibfield  {title} {\bibinfo {title} {Measurement of neutron quantum interference in noninertial frames},\ }\href {https://doi.org/10.1103/PhysRevLett.51.1401} {\bibfield  {journal} {\bibinfo  {journal} {Phys. Rev. Lett.}\ }\textbf {\bibinfo {volume} {51}},\ \bibinfo {pages} {1401} (\bibinfo {year} {1983})}\BibitemShut {NoStop}%
\bibitem [{\citenamefont {Abele}\ and\ \citenamefont {Leeb}(2012)}]{abele2012}%
  \BibitemOpen
  \bibfield  {author} {\bibinfo {author} {\bibfnamefont {H.}~\bibnamefont {Abele}}\ and\ \bibinfo {author} {\bibfnamefont {H.}~\bibnamefont {Leeb}},\ }\bibfield  {title} {\bibinfo {title} {Gravitation and quantum interference experiments with neutrons},\ }\href {https://doi.org/10.1088/1367-2630/14/5/055010} {\bibfield  {journal} {\bibinfo  {journal} {New Journal of Physics}\ }\textbf {\bibinfo {volume} {14}},\ \bibinfo {pages} {055010} (\bibinfo {year} {2012})}\BibitemShut {NoStop}%
\bibitem [{\citenamefont {Overstreet}\ \emph {et~al.}(2022)\citenamefont {Overstreet}, \citenamefont {Asenbaum}, \citenamefont {Curti}, \citenamefont {Kim},\ and\ \citenamefont {Kasevich}}]{kasevich2022}%
  \BibitemOpen
  \bibfield  {author} {\bibinfo {author} {\bibfnamefont {C.}~\bibnamefont {Overstreet}}, \bibinfo {author} {\bibfnamefont {P.}~\bibnamefont {Asenbaum}}, \bibinfo {author} {\bibfnamefont {J.}~\bibnamefont {Curti}}, \bibinfo {author} {\bibfnamefont {M.}~\bibnamefont {Kim}},\ and\ \bibinfo {author} {\bibfnamefont {M.~A.}\ \bibnamefont {Kasevich}},\ }\bibfield  {title} {\bibinfo {title} {Observation of a gravitational {Aharonov-Bohm} effect},\ }\href {https://doi.org/10.1126/science.abl7152} {\bibfield  {journal} {\bibinfo  {journal} {Science}\ }\textbf {\bibinfo {volume} {375}},\ \bibinfo {pages} {226} (\bibinfo {year} {2022})}\BibitemShut {NoStop}%
\bibitem [{\citenamefont {Overstreet}\ \emph {et~al.}(2023)\citenamefont {Overstreet}, \citenamefont {Curti}, \citenamefont {Kim}, \citenamefont {Asenbaum}, \citenamefont {Kasevich},\ and\ \citenamefont {Giacomini}}]{overstreet2023}%
  \BibitemOpen
  \bibfield  {author} {\bibinfo {author} {\bibfnamefont {C.}~\bibnamefont {Overstreet}}, \bibinfo {author} {\bibfnamefont {J.}~\bibnamefont {Curti}}, \bibinfo {author} {\bibfnamefont {M.}~\bibnamefont {Kim}}, \bibinfo {author} {\bibfnamefont {P.}~\bibnamefont {Asenbaum}}, \bibinfo {author} {\bibfnamefont {M.~A.}\ \bibnamefont {Kasevich}},\ and\ \bibinfo {author} {\bibfnamefont {F.}~\bibnamefont {Giacomini}},\ }\bibfield  {title} {\bibinfo {title} {Inference of gravitational field superposition from quantum measurements},\ }\href {https://doi.org/10.1103/PhysRevD.108.084038} {\bibfield  {journal} {\bibinfo  {journal} {Phys. Rev. D}\ }\textbf {\bibinfo {volume} {108}},\ \bibinfo {pages} {084038} (\bibinfo {year} {2023})}\BibitemShut {NoStop}%
\bibitem [{\citenamefont {Asenbaum}\ \emph {et~al.}(2017)\citenamefont {Asenbaum}, \citenamefont {Overstreet}, \citenamefont {Kovachy}, \citenamefont {Brown}, \citenamefont {Hogan},\ and\ \citenamefont {Kasevich}}]{asenbaum2017}%
  \BibitemOpen
  \bibfield  {author} {\bibinfo {author} {\bibfnamefont {P.}~\bibnamefont {Asenbaum}}, \bibinfo {author} {\bibfnamefont {C.}~\bibnamefont {Overstreet}}, \bibinfo {author} {\bibfnamefont {T.}~\bibnamefont {Kovachy}}, \bibinfo {author} {\bibfnamefont {D.~D.}\ \bibnamefont {Brown}}, \bibinfo {author} {\bibfnamefont {J.~M.}\ \bibnamefont {Hogan}},\ and\ \bibinfo {author} {\bibfnamefont {M.~A.}\ \bibnamefont {Kasevich}},\ }\bibfield  {title} {\bibinfo {title} {Phase shift in an atom interferometer due to spacetime curvature across its wave function},\ }\href {https://doi.org/10.1103/PhysRevLett.118.183602} {\bibfield  {journal} {\bibinfo  {journal} {Phys. Rev. Lett.}\ }\textbf {\bibinfo {volume} {118}},\ \bibinfo {pages} {183602} (\bibinfo {year} {2017})}\BibitemShut {NoStop}%
\bibitem [{\citenamefont {Bothwell}\ \emph {et~al.}(2022)\citenamefont {Bothwell}, \citenamefont {Kennedy}, \citenamefont {Aeppli}, \citenamefont {Kedar}, \citenamefont {Robinson}, \citenamefont {Oelker}, \citenamefont {Staron},\ and\ \citenamefont {Ye}}]{bothwell2022}%
  \BibitemOpen
  \bibfield  {author} {\bibinfo {author} {\bibfnamefont {T.}~\bibnamefont {Bothwell}}, \bibinfo {author} {\bibfnamefont {C.~J.}\ \bibnamefont {Kennedy}}, \bibinfo {author} {\bibfnamefont {A.}~\bibnamefont {Aeppli}}, \bibinfo {author} {\bibfnamefont {D.}~\bibnamefont {Kedar}}, \bibinfo {author} {\bibfnamefont {J.~M.}\ \bibnamefont {Robinson}}, \bibinfo {author} {\bibfnamefont {E.}~\bibnamefont {Oelker}}, \bibinfo {author} {\bibfnamefont {A.}~\bibnamefont {Staron}},\ and\ \bibinfo {author} {\bibfnamefont {J.}~\bibnamefont {Ye}},\ }\bibfield  {title} {\bibinfo {title} {Resolving the gravitational redshift across a millimetre-scale atomic sample},\ }\href {https://doi.org/10.1038/s41586-021-04349-7} {\bibfield  {journal} {\bibinfo  {journal} {Nature}\ }\textbf {\bibinfo {volume} {602}},\ \bibinfo {pages} {420} (\bibinfo {year} {2022})}\BibitemShut {NoStop}%
\bibitem [{\citenamefont {Xu}\ \emph {et~al.}(2019)\citenamefont {Xu}, \citenamefont {Jaffe}, \citenamefont {Panda}, \citenamefont {Kristensen}, \citenamefont {Clark},\ and\ \citenamefont {Müller}}]{xu2019}%
  \BibitemOpen
  \bibfield  {author} {\bibinfo {author} {\bibfnamefont {V.}~\bibnamefont {Xu}}, \bibinfo {author} {\bibfnamefont {M.}~\bibnamefont {Jaffe}}, \bibinfo {author} {\bibfnamefont {C.~D.}\ \bibnamefont {Panda}}, \bibinfo {author} {\bibfnamefont {S.~L.}\ \bibnamefont {Kristensen}}, \bibinfo {author} {\bibfnamefont {L.~W.}\ \bibnamefont {Clark}},\ and\ \bibinfo {author} {\bibfnamefont {H.}~\bibnamefont {Müller}},\ }\bibfield  {title} {\bibinfo {title} {Probing gravity by holding atoms for 20 seconds},\ }\href {https://doi.org/10.1126/science.aay6428} {\bibfield  {journal} {\bibinfo  {journal} {Science}\ }\textbf {\bibinfo {volume} {366}},\ \bibinfo {pages} {745} (\bibinfo {year} {2019})}\BibitemShut {NoStop}%
\bibitem [{\citenamefont {Müller}\ \emph {et~al.}(2010)\citenamefont {Müller}, \citenamefont {Peters},\ and\ \citenamefont {Chu}}]{mueller2010}%
  \BibitemOpen
  \bibfield  {author} {\bibinfo {author} {\bibfnamefont {H.}~\bibnamefont {Müller}}, \bibinfo {author} {\bibfnamefont {A.}~\bibnamefont {Peters}},\ and\ \bibinfo {author} {\bibfnamefont {S.}~\bibnamefont {Chu}},\ }\bibfield  {title} {\bibinfo {title} {A precision measurement of the gravitational redshift by the interference of matter waves},\ }\href {https://doi.org/https://doi.org/10.1038/nature08776} {\bibfield  {journal} {\bibinfo  {journal} {Nature}\ }\textbf {\bibinfo {volume} {463}},\ \bibinfo {pages} {926} (\bibinfo {year} {2010})}\BibitemShut {NoStop}%
\bibitem [{\citenamefont {Kiefer}\ and\ \citenamefont {Weber}(2005)}]{kiefer2005}%
  \BibitemOpen
  \bibfield  {author} {\bibinfo {author} {\bibfnamefont {C.}~\bibnamefont {Kiefer}}\ and\ \bibinfo {author} {\bibfnamefont {C.}~\bibnamefont {Weber}},\ }\bibfield  {title} {\bibinfo {title} {On the interaction of mesoscopic quantum systems with gravity},\ }\href {https://doi.org/https://doi.org/10.1002/andp.20055170404} {\bibfield  {journal} {\bibinfo  {journal} {Annalen der Physik}\ }\textbf {\bibinfo {volume} {517}},\ \bibinfo {pages} {253} (\bibinfo {year} {2005})}\BibitemShut {NoStop}%
\bibitem [{\citenamefont {Lanzagorta}(2014)}]{lanzagorta}%
  \BibitemOpen
  \bibfield  {author} {\bibinfo {author} {\bibfnamefont {M.}~\bibnamefont {Lanzagorta}},\ }\href {https://doi.org/10.1088/978-1-627-05330-3} {\emph {\bibinfo {title} {Quantum Information in Gravitational Fields}}}\ (\bibinfo  {publisher} {Morgan \& Claypool Publishers},\ \bibinfo {address} {San Rafael},\ \bibinfo {year} {2014})\BibitemShut {NoStop}%
\bibitem [{\citenamefont {Hehl}\ \emph {et~al.}(2004)\citenamefont {Hehl}, \citenamefont {Obukhov},\ and\ \citenamefont {Rosenow}}]{hehl2004}%
  \BibitemOpen
  \bibfield  {author} {\bibinfo {author} {\bibfnamefont {F.~W.}\ \bibnamefont {Hehl}}, \bibinfo {author} {\bibfnamefont {Y.~N.}\ \bibnamefont {Obukhov}},\ and\ \bibinfo {author} {\bibfnamefont {B.}~\bibnamefont {Rosenow}},\ }\bibfield  {title} {\bibinfo {title} {Is the quantum {Hall} effect influenced by the gravitational field?},\ }\href {https://doi.org/10.1103/PhysRevLett.93.096804} {\bibfield  {journal} {\bibinfo  {journal} {Phys. Rev. Lett.}\ }\textbf {\bibinfo {volume} {93}},\ \bibinfo {pages} {096804} (\bibinfo {year} {2004})}\BibitemShut {NoStop}%
\bibitem [{\citenamefont {Hammad}\ \emph {et~al.}(2020)\citenamefont {Hammad}, \citenamefont {Landry},\ and\ \citenamefont {Mathieu}}]{hammad2020}%
  \BibitemOpen
  \bibfield  {author} {\bibinfo {author} {\bibfnamefont {F.}~\bibnamefont {Hammad}}, \bibinfo {author} {\bibfnamefont {A.}~\bibnamefont {Landry}},\ and\ \bibinfo {author} {\bibfnamefont {K.}~\bibnamefont {Mathieu}},\ }\bibfield  {title} {\bibinfo {title} {A fresh look at the influence of gravity on the quantum {Hall} effect},\ }\href {https://doi.org/https://doi.org/10.1140/epjp/s13360-020-00481-x} {\bibfield  {journal} {\bibinfo  {journal} {Eur. Phys. J. Plus}\ }\textbf {\bibinfo {volume} {135}},\ \bibinfo {pages} {449} (\bibinfo {year} {2020})}\BibitemShut {NoStop}%
\bibitem [{\citenamefont {Brooks}(2023)}]{brooks2023}%
  \BibitemOpen
  \bibfield  {author} {\bibinfo {author} {\bibfnamefont {M.}~\bibnamefont {Brooks}},\ }\bibfield  {title} {\bibinfo {title} {Quantum computers: what are they good for?},\ }\href {https://doi.org/10.1038/d41586-023-01692-9} {\bibfield  {journal} {\bibinfo  {journal} {Nature}\ }\textbf {\bibinfo {volume} {617}},\ \bibinfo {pages} {S1–S3} (\bibinfo {year} {2023})}\BibitemShut {NoStop}%
\bibitem [{\citenamefont {Krantz}\ \emph {et~al.}(2019)\citenamefont {Krantz}, \citenamefont {Kjaergaard}, \citenamefont {Yan}, \citenamefont {Orlando}, \citenamefont {Gustavsson},\ and\ \citenamefont {Oliver}}]{krantz2019}%
  \BibitemOpen
  \bibfield  {author} {\bibinfo {author} {\bibfnamefont {P.}~\bibnamefont {Krantz}}, \bibinfo {author} {\bibfnamefont {M.}~\bibnamefont {Kjaergaard}}, \bibinfo {author} {\bibfnamefont {F.}~\bibnamefont {Yan}}, \bibinfo {author} {\bibfnamefont {T.~P.}\ \bibnamefont {Orlando}}, \bibinfo {author} {\bibfnamefont {S.}~\bibnamefont {Gustavsson}},\ and\ \bibinfo {author} {\bibfnamefont {W.~D.}\ \bibnamefont {Oliver}},\ }\bibfield  {title} {\bibinfo {title} {{A quantum engineer's guide to superconducting qubits}},\ }\href {https://doi.org/10.1063/1.5089550} {\bibfield  {journal} {\bibinfo  {journal} {App. Phys. Rev.}\ }\textbf {\bibinfo {volume} {6}},\ \bibinfo {pages} {021318} (\bibinfo {year} {2019})}\BibitemShut {NoStop}%
\bibitem [{\citenamefont {Pound}\ and\ \citenamefont {Rebka}(1960)}]{poundrebka}%
  \BibitemOpen
  \bibfield  {author} {\bibinfo {author} {\bibfnamefont {R.~V.}\ \bibnamefont {Pound}}\ and\ \bibinfo {author} {\bibfnamefont {G.~A.}\ \bibnamefont {Rebka}},\ }\bibfield  {title} {\bibinfo {title} {Apparent weight of photons},\ }\href {https://doi.org/10.1103/PhysRevLett.4.337} {\bibfield  {journal} {\bibinfo  {journal} {Phys. Rev. Lett.}\ }\textbf {\bibinfo {volume} {4}},\ \bibinfo {pages} {337} (\bibinfo {year} {1960})}\BibitemShut {NoStop}%
\bibitem [{\citenamefont {Hafele}\ and\ \citenamefont {Keating}(1977)}]{hafelekeating}%
  \BibitemOpen
  \bibfield  {author} {\bibinfo {author} {\bibfnamefont {J.~C.}\ \bibnamefont {Hafele}}\ and\ \bibinfo {author} {\bibfnamefont {R.}~\bibnamefont {Keating}},\ }\bibfield  {title} {\bibinfo {title} {Around-the-world atomic clocks: predicted relativistic time gains},\ }\href {https://doi.org/https://doi.org/10.1126/science.177.4044.168} {\bibfield  {journal} {\bibinfo  {journal} {Science}\ }\textbf {\bibinfo {volume} {177}},\ \bibinfo {pages} {166} (\bibinfo {year} {1977})}\BibitemShut {NoStop}%
\bibitem [{\citenamefont {Zych}\ \emph {et~al.}(2011)\citenamefont {Zych}, \citenamefont {Costa}, \citenamefont {Pikovski},\ and\ \citenamefont {Časlav Brukner}}]{zych2011}%
  \BibitemOpen
  \bibfield  {author} {\bibinfo {author} {\bibfnamefont {M.}~\bibnamefont {Zych}}, \bibinfo {author} {\bibfnamefont {F.}~\bibnamefont {Costa}}, \bibinfo {author} {\bibfnamefont {I.}~\bibnamefont {Pikovski}},\ and\ \bibinfo {author} {\bibnamefont {Časlav Brukner}},\ }\bibfield  {title} {\bibinfo {title} {Quantum interferometric visibility as a witness of general relativistic proper time},\ }\href {https://doi.org/10.1038/ncomms1498} {\bibfield  {journal} {\bibinfo  {journal} {Nat. Commun.}\ }\textbf {\bibinfo {volume} {2}},\ \bibinfo {pages} {505} (\bibinfo {year} {2011})}\BibitemShut {NoStop}%
\bibitem [{\citenamefont {Pikovski}\ \emph {et~al.}(2015)\citenamefont {Pikovski}, \citenamefont {Zych}, \citenamefont {Costa},\ and\ \citenamefont {Bruckner}}]{pikovski2015}%
  \BibitemOpen
  \bibfield  {author} {\bibinfo {author} {\bibfnamefont {I.}~\bibnamefont {Pikovski}}, \bibinfo {author} {\bibfnamefont {M.}~\bibnamefont {Zych}}, \bibinfo {author} {\bibfnamefont {F.}~\bibnamefont {Costa}},\ and\ \bibinfo {author} {\bibfnamefont {{\v{C}}.}~\bibnamefont {Bruckner}},\ }\bibfield  {title} {\bibinfo {title} {Universal decoherence due to gravitational time dilation},\ }\href {https://doi.org/10.1038/nphys3366} {\bibfield  {journal} {\bibinfo  {journal} {Nature Phys.}\ }\textbf {\bibinfo {volume} {11}},\ \bibinfo {pages} {668–672} (\bibinfo {year} {2015})}\BibitemShut {NoStop}%
\bibitem [{\citenamefont {Bongs}\ \emph {et~al.}(2019)\citenamefont {Bongs}, \citenamefont {Holynski}, \citenamefont {Vovrosh}, \citenamefont {Bouyer}, \citenamefont {Condon}, \citenamefont {Rasel}, \citenamefont {Schubert}, \citenamefont {Schleich},\ and\ \citenamefont {Roura}}]{bongs2019}%
  \BibitemOpen
  \bibfield  {author} {\bibinfo {author} {\bibfnamefont {K.}~\bibnamefont {Bongs}}, \bibinfo {author} {\bibfnamefont {M.}~\bibnamefont {Holynski}}, \bibinfo {author} {\bibfnamefont {J.}~\bibnamefont {Vovrosh}}, \bibinfo {author} {\bibfnamefont {P.}~\bibnamefont {Bouyer}}, \bibinfo {author} {\bibfnamefont {G.}~\bibnamefont {Condon}}, \bibinfo {author} {\bibfnamefont {E.}~\bibnamefont {Rasel}}, \bibinfo {author} {\bibfnamefont {C.}~\bibnamefont {Schubert}}, \bibinfo {author} {\bibfnamefont {W.~P.}\ \bibnamefont {Schleich}},\ and\ \bibinfo {author} {\bibfnamefont {A.}~\bibnamefont {Roura}},\ }\bibfield  {title} {\bibinfo {title} {Taking atom interferometric quantum sensors from the laboratory to real-world applications},\ }\href {https://doi.org/https://doi.org/10.1038/s42254-019-0117-4} {\bibfield  {journal} {\bibinfo  {journal} {Nat. Rev. Phys.}\ }\textbf {\bibinfo {volume} {1}},\ \bibinfo {pages} {731} (\bibinfo {year} {2019})}\BibitemShut {NoStop}%
\bibitem [{\citenamefont {Stray}\ \emph {et~al.}(2022)\citenamefont {Stray}, \citenamefont {Lamb}, \citenamefont {Kaushik}, \citenamefont {Vovrosh}, \citenamefont {Rodgers}, \citenamefont {Winch}, \citenamefont {Hayati}, \citenamefont {Boddice}, \citenamefont {Stabrawa}, \citenamefont {Niggebaum}, \citenamefont {Langlois}, \citenamefont {Lien}, \citenamefont {Lellouch}, \citenamefont {Roshanmanesh}, \citenamefont {Ridley}, \citenamefont {{de~Villiers}}, \citenamefont {Brown}, \citenamefont {Cross}, \citenamefont {Tuckwell}, \citenamefont {Faramarzi}, \citenamefont {Metje}, \citenamefont {Bongs},\ and\ \citenamefont {Holynski}}]{stray2022}%
  \BibitemOpen
  \bibfield  {author} {\bibinfo {author} {\bibfnamefont {B.}~\bibnamefont {Stray}}, \bibinfo {author} {\bibfnamefont {A.}~\bibnamefont {Lamb}}, \bibinfo {author} {\bibfnamefont {A.}~\bibnamefont {Kaushik}}, \bibinfo {author} {\bibfnamefont {J.}~\bibnamefont {Vovrosh}}, \bibinfo {author} {\bibfnamefont {A.}~\bibnamefont {Rodgers}}, \bibinfo {author} {\bibfnamefont {J.}~\bibnamefont {Winch}}, \bibinfo {author} {\bibfnamefont {F.}~\bibnamefont {Hayati}}, \bibinfo {author} {\bibfnamefont {D.}~\bibnamefont {Boddice}}, \bibinfo {author} {\bibfnamefont {A.}~\bibnamefont {Stabrawa}}, \bibinfo {author} {\bibfnamefont {A.}~\bibnamefont {Niggebaum}}, \bibinfo {author} {\bibfnamefont {M.}~\bibnamefont {Langlois}}, \bibinfo {author} {\bibfnamefont {Y.-H.}\ \bibnamefont {Lien}}, \bibinfo {author} {\bibfnamefont {S.}~\bibnamefont {Lellouch}}, \bibinfo {author} {\bibfnamefont {S.}~\bibnamefont {Roshanmanesh}}, \bibinfo {author} {\bibfnamefont {K.}~\bibnamefont {Ridley}}, \bibinfo {author} {\bibfnamefont {G.}~\bibnamefont
  {{de~Villiers}}}, \bibinfo {author} {\bibfnamefont {G.}~\bibnamefont {Brown}}, \bibinfo {author} {\bibfnamefont {T.}~\bibnamefont {Cross}}, \bibinfo {author} {\bibfnamefont {G.}~\bibnamefont {Tuckwell}}, \bibinfo {author} {\bibfnamefont {A.}~\bibnamefont {Faramarzi}}, \bibinfo {author} {\bibfnamefont {N.}~\bibnamefont {Metje}}, \bibinfo {author} {\bibfnamefont {K.}~\bibnamefont {Bongs}},\ and\ \bibinfo {author} {\bibfnamefont {M.}~\bibnamefont {Holynski}},\ }\bibfield  {title} {\bibinfo {title} {Quantum sensing for gravity cartography},\ }\href {https://doi.org/10.1038/s41586-021-04315-3} {\bibfield  {journal} {\bibinfo  {journal} {Nature}\ }\textbf {\bibinfo {volume} {602}},\ \bibinfo {pages} {590} (\bibinfo {year} {2022})}\BibitemShut {NoStop}%
\bibitem [{\citenamefont {Afanasiev}\ \emph {et~al.}(2024)\citenamefont {Afanasiev}, \citenamefont {Skakunenko},\ and\ \citenamefont {Balykin}}]{afanasiev2024}%
  \BibitemOpen
  \bibfield  {author} {\bibinfo {author} {\bibfnamefont {A.}~\bibnamefont {Afanasiev}}, \bibinfo {author} {\bibfnamefont {P.}~\bibnamefont {Skakunenko}},\ and\ \bibinfo {author} {\bibfnamefont {V.}~\bibnamefont {Balykin}},\ }\bibfield  {title} {\bibinfo {title} {Cold atom gravimeter based on an atomic fountain and a microwave transition},\ }\href {https://doi.org/https://doi.org/10.1134/S002136402360372X} {\bibfield  {journal} {\bibinfo  {journal} {Jetp Lett}\ }\textbf {\bibinfo {volume} {119}},\ \bibinfo {pages} {84} (\bibinfo {year} {2024})}\BibitemShut {NoStop}%
\bibitem [{\citenamefont {Bongs}\ \emph {et~al.}(2023)\citenamefont {Bongs}, \citenamefont {Bennett},\ and\ \citenamefont {Lohmann}}]{bongs2023}%
  \BibitemOpen
  \bibfield  {author} {\bibinfo {author} {\bibfnamefont {K.}~\bibnamefont {Bongs}}, \bibinfo {author} {\bibfnamefont {S.}~\bibnamefont {Bennett}},\ and\ \bibinfo {author} {\bibfnamefont {A.}~\bibnamefont {Lohmann}},\ }\bibfield  {title} {\bibinfo {title} {Quantum sensors will start a revolution -- if we deploy them right},\ }\href {https://doi.org/https://doi.org/10.1038/d41586-023-01663-0} {\bibfield  {journal} {\bibinfo  {journal} {Nature}\ }\textbf {\bibinfo {volume} {617}},\ \bibinfo {pages} {672} (\bibinfo {year} {2023})}\BibitemShut {NoStop}%
\bibitem [{\citenamefont {Werner}\ \emph {et~al.}(1979)\citenamefont {Werner}, \citenamefont {Staudenmann},\ and\ \citenamefont {Colella}}]{werner1979}%
  \BibitemOpen
  \bibfield  {author} {\bibinfo {author} {\bibfnamefont {S.~A.}\ \bibnamefont {Werner}}, \bibinfo {author} {\bibfnamefont {J.~L.}\ \bibnamefont {Staudenmann}},\ and\ \bibinfo {author} {\bibfnamefont {R.}~\bibnamefont {Colella}},\ }\bibfield  {title} {\bibinfo {title} {Effect of {Earth's} rotation on the quantum mechanical phase of the neutron},\ }\href {https://doi.org/10.1103/PhysRevLett.42.1103} {\bibfield  {journal} {\bibinfo  {journal} {Phys. Rev. Lett.}\ }\textbf {\bibinfo {volume} {42}},\ \bibinfo {pages} {1103} (\bibinfo {year} {1979})}\BibitemShut {NoStop}%
\bibitem [{\citenamefont {Zheng}\ \emph {et~al.}(2022)\citenamefont {Zheng}, \citenamefont {Dolde}, \citenamefont {Lochab}, \citenamefont {Merriman}, \citenamefont {Li},\ and\ \citenamefont {Kolkowitz}}]{zheng2022}%
  \BibitemOpen
  \bibfield  {author} {\bibinfo {author} {\bibfnamefont {X.}~\bibnamefont {Zheng}}, \bibinfo {author} {\bibfnamefont {J.}~\bibnamefont {Dolde}}, \bibinfo {author} {\bibfnamefont {V.}~\bibnamefont {Lochab}}, \bibinfo {author} {\bibfnamefont {B.~N.}\ \bibnamefont {Merriman}}, \bibinfo {author} {\bibfnamefont {H.}~\bibnamefont {Li}},\ and\ \bibinfo {author} {\bibfnamefont {S.}~\bibnamefont {Kolkowitz}},\ }\bibfield  {title} {\bibinfo {title} {Differential clock comparisons with a multiplexed optical lattice clock},\ }\href {https://doi.org/10.1038/s41586-021-04344-y} {\bibfield  {journal} {\bibinfo  {journal} {Nature}\ }\textbf {\bibinfo {volume} {602}},\ \bibinfo {pages} {425} (\bibinfo {year} {2022})}\BibitemShut {NoStop}%
\bibitem [{\citenamefont {Pikovski}\ \emph {et~al.}(2017)\citenamefont {Pikovski}, \citenamefont {Zych}, \citenamefont {Costa},\ and\ \citenamefont {Bruckner}}]{pikovski2017}%
  \BibitemOpen
  \bibfield  {author} {\bibinfo {author} {\bibfnamefont {I.}~\bibnamefont {Pikovski}}, \bibinfo {author} {\bibfnamefont {M.}~\bibnamefont {Zych}}, \bibinfo {author} {\bibfnamefont {F.}~\bibnamefont {Costa}},\ and\ \bibinfo {author} {\bibfnamefont {{\v{C}}.}~\bibnamefont {Bruckner}},\ }\bibfield  {title} {\bibinfo {title} {Time dilation in quantum systems and decoherence},\ }\href {https://doi.org/10.1088/1367-2630/aa5d92} {\bibfield  {journal} {\bibinfo  {journal} {New. J. Phys.}\ }\textbf {\bibinfo {volume} {19}},\ \bibinfo {pages} {025011} (\bibinfo {year} {2017})}\BibitemShut {NoStop}%
\bibitem [{\citenamefont {Aharonov}\ and\ \citenamefont {Bohm}(1959)}]{PhysRev.115.485}%
  \BibitemOpen
  \bibfield  {author} {\bibinfo {author} {\bibfnamefont {Y.}~\bibnamefont {Aharonov}}\ and\ \bibinfo {author} {\bibfnamefont {D.}~\bibnamefont {Bohm}},\ }\bibfield  {title} {\bibinfo {title} {Significance of electromagnetic potentials in the quantum theory},\ }\href {https://doi.org/10.1103/PhysRev.115.485} {\bibfield  {journal} {\bibinfo  {journal} {Phys. Rev.}\ }\textbf {\bibinfo {volume} {115}},\ \bibinfo {pages} {485} (\bibinfo {year} {1959})}\BibitemShut {NoStop}%
\bibitem [{\citenamefont {Kraemer}\ \emph {et~al.}(2023)\citenamefont {Kraemer}, \citenamefont {Janni~Moens}, \citenamefont {Bara}, \citenamefont {Beeks}, \citenamefont {Chhetri}, \citenamefont {Chrysalidis}, \citenamefont {Claessens}, \citenamefont {Cocolios}, \citenamefont {Correia}, \citenamefont {Witte}, \citenamefont {Ferrer}, \citenamefont {Geldhof}, \citenamefont {Heinke}, \citenamefont {Hosseini}, \citenamefont {Huyse}, \citenamefont {Köster}, \citenamefont {Kudryavtsev}, \citenamefont {Laatiaoui}, \citenamefont {Lica}, \citenamefont {Magchiels}, \citenamefont {Manea}, \citenamefont {Merckling}, \citenamefont {Pereira}, \citenamefont {Raeder}, \citenamefont {Schumm}, \citenamefont {Sels}, \citenamefont {Thirolf}, \citenamefont {Tunhuma}, \citenamefont {Bergh}, \citenamefont {Duppen}, \citenamefont {Vantomme}, \citenamefont {Verlinde}, \citenamefont {Villarreal},\ and\ \citenamefont {Wahl}}]{kraemer2023}%
  \BibitemOpen
  \bibfield  {author} {\bibinfo {author} {\bibfnamefont {S.}~\bibnamefont {Kraemer}}, \bibinfo {author} {\bibfnamefont {M.~A.-K.}\ \bibnamefont {Janni~Moens}}, \bibinfo {author} {\bibfnamefont {S.}~\bibnamefont {Bara}}, \bibinfo {author} {\bibfnamefont {K.}~\bibnamefont {Beeks}}, \bibinfo {author} {\bibfnamefont {P.}~\bibnamefont {Chhetri}}, \bibinfo {author} {\bibfnamefont {K.}~\bibnamefont {Chrysalidis}}, \bibinfo {author} {\bibfnamefont {A.}~\bibnamefont {Claessens}}, \bibinfo {author} {\bibfnamefont {T.~E.}\ \bibnamefont {Cocolios}}, \bibinfo {author} {\bibfnamefont {J.~G.~M.}\ \bibnamefont {Correia}}, \bibinfo {author} {\bibfnamefont {H.~D.}\ \bibnamefont {Witte}}, \bibinfo {author} {\bibfnamefont {R.}~\bibnamefont {Ferrer}}, \bibinfo {author} {\bibfnamefont {S.}~\bibnamefont {Geldhof}}, \bibinfo {author} {\bibfnamefont {R.}~\bibnamefont {Heinke}}, \bibinfo {author} {\bibfnamefont {N.}~\bibnamefont {Hosseini}}, \bibinfo {author} {\bibfnamefont {M.}~\bibnamefont {Huyse}}, \bibinfo {author} {\bibfnamefont
  {U.}~\bibnamefont {Köster}}, \bibinfo {author} {\bibfnamefont {Y.}~\bibnamefont {Kudryavtsev}}, \bibinfo {author} {\bibfnamefont {M.}~\bibnamefont {Laatiaoui}}, \bibinfo {author} {\bibfnamefont {R.}~\bibnamefont {Lica}}, \bibinfo {author} {\bibfnamefont {G.}~\bibnamefont {Magchiels}}, \bibinfo {author} {\bibfnamefont {V.}~\bibnamefont {Manea}}, \bibinfo {author} {\bibfnamefont {C.}~\bibnamefont {Merckling}}, \bibinfo {author} {\bibfnamefont {L.~M.~C.}\ \bibnamefont {Pereira}}, \bibinfo {author} {\bibfnamefont {S.}~\bibnamefont {Raeder}}, \bibinfo {author} {\bibfnamefont {T.}~\bibnamefont {Schumm}}, \bibinfo {author} {\bibfnamefont {S.}~\bibnamefont {Sels}}, \bibinfo {author} {\bibfnamefont {P.~G.}\ \bibnamefont {Thirolf}}, \bibinfo {author} {\bibfnamefont {S.~M.}\ \bibnamefont {Tunhuma}}, \bibinfo {author} {\bibfnamefont {P.~V.~D.}\ \bibnamefont {Bergh}}, \bibinfo {author} {\bibfnamefont {P.~V.}\ \bibnamefont {Duppen}}, \bibinfo {author} {\bibfnamefont {A.}~\bibnamefont {Vantomme}}, \bibinfo {author}
  {\bibfnamefont {M.}~\bibnamefont {Verlinde}}, \bibinfo {author} {\bibfnamefont {R.}~\bibnamefont {Villarreal}},\ and\ \bibinfo {author} {\bibfnamefont {U.}~\bibnamefont {Wahl}},\ }\bibfield  {title} {\bibinfo {title} {Observation of the radiative decay of the $^{229}${Th} nuclear clock isomer},\ }\href {https://doi.org/https://doi.org/10.1038/s41586-023-05894-z} {\bibfield  {journal} {\bibinfo  {journal} {Nature}\ }\textbf {\bibinfo {volume} {617}},\ \bibinfo {pages} {706} (\bibinfo {year} {2023})}\BibitemShut {NoStop}%
\bibitem [{\citenamefont {Tiedau}\ \emph {et~al.}(2024)\citenamefont {Tiedau}, \citenamefont {Okhapkin}, \citenamefont {Zhang}, \citenamefont {Thielking}, \citenamefont {Zitzer}, \citenamefont {Peik}, \citenamefont {Schaden}, \citenamefont {Pronebner}, \citenamefont {Morawetz}, \citenamefont {De~Col}, \citenamefont {Schneider}, \citenamefont {Leitner}, \citenamefont {Pressler}, \citenamefont {Kazakov}, \citenamefont {Beeks}, \citenamefont {Sikorsky},\ and\ \citenamefont {Schumm}}]{tiedau2024}%
  \BibitemOpen
  \bibfield  {author} {\bibinfo {author} {\bibfnamefont {J.}~\bibnamefont {Tiedau}}, \bibinfo {author} {\bibfnamefont {M.~V.}\ \bibnamefont {Okhapkin}}, \bibinfo {author} {\bibfnamefont {K.}~\bibnamefont {Zhang}}, \bibinfo {author} {\bibfnamefont {J.}~\bibnamefont {Thielking}}, \bibinfo {author} {\bibfnamefont {G.}~\bibnamefont {Zitzer}}, \bibinfo {author} {\bibfnamefont {E.}~\bibnamefont {Peik}}, \bibinfo {author} {\bibfnamefont {F.}~\bibnamefont {Schaden}}, \bibinfo {author} {\bibfnamefont {T.}~\bibnamefont {Pronebner}}, \bibinfo {author} {\bibfnamefont {I.}~\bibnamefont {Morawetz}}, \bibinfo {author} {\bibfnamefont {L.~T.}\ \bibnamefont {De~Col}}, \bibinfo {author} {\bibfnamefont {F.}~\bibnamefont {Schneider}}, \bibinfo {author} {\bibfnamefont {A.}~\bibnamefont {Leitner}}, \bibinfo {author} {\bibfnamefont {M.}~\bibnamefont {Pressler}}, \bibinfo {author} {\bibfnamefont {G.~A.}\ \bibnamefont {Kazakov}}, \bibinfo {author} {\bibfnamefont {K.}~\bibnamefont {Beeks}}, \bibinfo {author} {\bibfnamefont
  {T.}~\bibnamefont {Sikorsky}},\ and\ \bibinfo {author} {\bibfnamefont {T.}~\bibnamefont {Schumm}},\ }\bibfield  {title} {\bibinfo {title} {Laser excitation of the {Th}-229 nucleus},\ }\href {https://doi.org/10.1103/PhysRevLett.132.182501} {\bibfield  {journal} {\bibinfo  {journal} {Phys. Rev. Lett.}\ }\textbf {\bibinfo {volume} {132}},\ \bibinfo {pages} {182501} (\bibinfo {year} {2024})}\BibitemShut {NoStop}%
\bibitem [{\citenamefont {Unruh}(1976)}]{unruh1976}%
  \BibitemOpen
  \bibfield  {author} {\bibinfo {author} {\bibfnamefont {W.~G.}\ \bibnamefont {Unruh}},\ }\bibfield  {title} {\bibinfo {title} {Notes on black-hole evaporation},\ }\href {https://doi.org/10.1103/PhysRevD.14.870} {\bibfield  {journal} {\bibinfo  {journal} {Phys. Rev. D}\ }\textbf {\bibinfo {volume} {14}},\ \bibinfo {pages} {870} (\bibinfo {year} {1976})}\BibitemShut {NoStop}%
\bibitem [{\citenamefont {Alsing}\ and\ \citenamefont {Milonni}(2004)}]{alsing2004}%
  \BibitemOpen
  \bibfield  {author} {\bibinfo {author} {\bibfnamefont {P.~M.}\ \bibnamefont {Alsing}}\ and\ \bibinfo {author} {\bibfnamefont {P.~W.}\ \bibnamefont {Milonni}},\ }\bibfield  {title} {\bibinfo {title} {Simplified derivation of the {Hawking}–{Unruh} temperature for an accelerated observer in vacuum},\ }\href {https://doi.org/https://doi.org/10.1119/1.1761064} {\bibfield  {journal} {\bibinfo  {journal} {Am. J. Phys.}\ }\textbf {\bibinfo {volume} {72}},\ \bibinfo {pages} {1524} (\bibinfo {year} {2004})}\BibitemShut {NoStop}%
\bibitem [{\citenamefont {Greenberger}\ \emph {et~al.}(1989)\citenamefont {Greenberger}, \citenamefont {Horne},\ and\ \citenamefont {Zeilinger}}]{greenberger1989}%
  \BibitemOpen
  \bibfield  {author} {\bibinfo {author} {\bibfnamefont {D.~M.}\ \bibnamefont {Greenberger}}, \bibinfo {author} {\bibfnamefont {M.~A.}\ \bibnamefont {Horne}},\ and\ \bibinfo {author} {\bibfnamefont {A.}~\bibnamefont {Zeilinger}},\ }\bibfield  {title} {\bibinfo {title} {{Going Beyond Bell\textquoteright{}s Theorem}},\ }\href {https://doi.org/10.1007/978-94-017-0849-4_10} {\bibfield  {journal} {\bibinfo  {journal} {Fundam. Theor. Phys.}\ }\textbf {\bibinfo {volume} {37}},\ \bibinfo {pages} {69} (\bibinfo {year} {1989})}\BibitemShut {NoStop}%
\bibitem [{\citenamefont {Mohammadbagherpoor}\ \emph {et~al.}(2019)\citenamefont {Mohammadbagherpoor}, \citenamefont {Oh}, \citenamefont {Dreher}, \citenamefont {Singh}, \citenamefont {Yu},\ and\ \citenamefont {Rindos}}]{mohammadbagherpoor2019}%
  \BibitemOpen
  \bibfield  {author} {\bibinfo {author} {\bibfnamefont {H.}~\bibnamefont {Mohammadbagherpoor}}, \bibinfo {author} {\bibfnamefont {Y.-H.}\ \bibnamefont {Oh}}, \bibinfo {author} {\bibfnamefont {P.}~\bibnamefont {Dreher}}, \bibinfo {author} {\bibfnamefont {A.}~\bibnamefont {Singh}}, \bibinfo {author} {\bibfnamefont {X.}~\bibnamefont {Yu}},\ and\ \bibinfo {author} {\bibfnamefont {A.~J.}\ \bibnamefont {Rindos}},\ }\href@noop {} {\bibinfo {title} {An improved implementation approach for quantum phase estimation on quantum computers}} (\bibinfo {year} {2019}),\ \Eprint {https://arxiv.org/abs/1910.11696} {arXiv:1910.11696 [quant-ph]} \BibitemShut {NoStop}%
\bibitem [{\citenamefont {Kitaev}\ \emph {et~al.}(2002)\citenamefont {Kitaev}, \citenamefont {Shen},\ and\ \citenamefont {Vyalyi}}]{Kitaev2002}%
  \BibitemOpen
  \bibfield  {author} {\bibinfo {author} {\bibfnamefont {A.}~\bibnamefont {Kitaev}}, \bibinfo {author} {\bibfnamefont {A.}~\bibnamefont {Shen}},\ and\ \bibinfo {author} {\bibfnamefont {M.}~\bibnamefont {Vyalyi}},\ }\href {https://doi.org/10.1090/gsm/047} {\emph {\bibinfo {title} {Classical and Quantum Computation}}}\ (\bibinfo  {publisher} {American Mathematical Society},\ \bibinfo {address} {Providence},\ \bibinfo {year} {2002})\BibitemShut {NoStop}%
\bibitem [{\citenamefont {Kitaev}(1995)}]{kitaev1995}%
  \BibitemOpen
  \bibfield  {author} {\bibinfo {author} {\bibfnamefont {A.~{\relax Yu}.}\ \bibnamefont {Kitaev}},\ }\href@noop {} {\bibinfo {title} {Quantum measurements and the {Abelian} stabilizer problem}} (\bibinfo {year} {1995}),\ \Eprint {https://arxiv.org/abs/quant-ph/9511026} {arXiv:quant-ph/9511026 [quant-ph]} \BibitemShut {NoStop}%
\bibitem [{\citenamefont {{Scientrex, CG-6 Autograv Gravity Meter}}()}]{scintrex}%
  \BibitemOpen
  \bibfield  {author} {\bibinfo {author} {\bibnamefont {{Scientrex, CG-6 Autograv Gravity Meter}}},\ }\href {https://scintrexltd.com/product/cg-6-autograv-gravity-meter/} {\bibinfo {title} {https://scintrexltd.com/product/cg-6-autograv-gravity-meter/}}\BibitemShut {NoStop}%
\bibitem [{\citenamefont {Ramírez}\ \emph {et~al.}(2020)\citenamefont {Ramírez}, \citenamefont {Urbina}, \citenamefont {Kleinschmidt}, \citenamefont {Finn}, \citenamefont {Edmunds}, \citenamefont {Esparza},\ and\ \citenamefont {Lipomi}}]{ramirez2020}%
  \BibitemOpen
  \bibfield  {author} {\bibinfo {author} {\bibfnamefont {J.}~\bibnamefont {Ramírez}}, \bibinfo {author} {\bibfnamefont {A.~D.}\ \bibnamefont {Urbina}}, \bibinfo {author} {\bibfnamefont {A.~T.}\ \bibnamefont {Kleinschmidt}}, \bibinfo {author} {\bibfnamefont {M.}~\bibnamefont {Finn}}, \bibinfo {author} {\bibfnamefont {S.~J.}\ \bibnamefont {Edmunds}}, \bibinfo {author} {\bibfnamefont {G.~L.}\ \bibnamefont {Esparza}},\ and\ \bibinfo {author} {\bibfnamefont {D.~J.}\ \bibnamefont {Lipomi}},\ }\bibfield  {title} {\bibinfo {title} {Exploring the limits of sensitivity for strain gauges of graphene and hexagonal boron nitride decorated with metallic nanoislands},\ }\href {https://doi.org/10.1039/D0NR02270E} {\bibfield  {journal} {\bibinfo  {journal} {Nanoscale}\ }\textbf {\bibinfo {volume} {12}},\ \bibinfo {pages} {11209} (\bibinfo {year} {2020})}\BibitemShut {NoStop}%
\bibitem [{\citenamefont {Cepollaro}\ \emph {et~al.}(2023)\citenamefont {Cepollaro}, \citenamefont {Giacomini},\ and\ \citenamefont {Paris}}]{cepollaro2023}%
  \BibitemOpen
  \bibfield  {author} {\bibinfo {author} {\bibfnamefont {C.}~\bibnamefont {Cepollaro}}, \bibinfo {author} {\bibfnamefont {F.}~\bibnamefont {Giacomini}},\ and\ \bibinfo {author} {\bibfnamefont {M.~G.~A.}\ \bibnamefont {Paris}},\ }\bibfield  {title} {\bibinfo {title} {Gravitational time dilation as a resource in quantum sensing},\ }\href {https://doi.org/10.22331/q-2023-03-13-946} {\bibfield  {journal} {\bibinfo  {journal} {Quantum}\ }\textbf {\bibinfo {volume} {7}},\ \bibinfo {pages} {946} (\bibinfo {year} {2023})}\BibitemShut {NoStop}%
\bibitem [{\citenamefont {Storz}\ \emph {et~al.}(2023)\citenamefont {Storz}, \citenamefont {Sch\"{a}r}, \citenamefont {Kulikov}, \citenamefont {Magnard}, \citenamefont {Kurpiers}, \citenamefont {L\"{u}tolf}, \citenamefont {Walter}, \citenamefont {Copetudo}, \citenamefont {Reuer}, \citenamefont {Akin}, \citenamefont {Besse}, \citenamefont {Gabureac}, \citenamefont {Norris}, \citenamefont {Rosario}, \citenamefont {Martin}, \citenamefont {Martinez}, \citenamefont {Amaya}, \citenamefont {Mitchell}, \citenamefont {Abellan}, \citenamefont {Bancal}, \citenamefont {Sangouard}, \citenamefont {Royer}, \citenamefont {Blais},\ and\ \citenamefont {Wallraff}}]{storz2023}%
  \BibitemOpen
  \bibfield  {author} {\bibinfo {author} {\bibfnamefont {S.}~\bibnamefont {Storz}}, \bibinfo {author} {\bibfnamefont {J.}~\bibnamefont {Sch\"{a}r}}, \bibinfo {author} {\bibfnamefont {A.}~\bibnamefont {Kulikov}}, \bibinfo {author} {\bibfnamefont {P.}~\bibnamefont {Magnard}}, \bibinfo {author} {\bibfnamefont {P.}~\bibnamefont {Kurpiers}}, \bibinfo {author} {\bibfnamefont {J.}~\bibnamefont {L\"{u}tolf}}, \bibinfo {author} {\bibfnamefont {T.}~\bibnamefont {Walter}}, \bibinfo {author} {\bibfnamefont {A.}~\bibnamefont {Copetudo}}, \bibinfo {author} {\bibfnamefont {K.}~\bibnamefont {Reuer}}, \bibinfo {author} {\bibfnamefont {A.}~\bibnamefont {Akin}}, \bibinfo {author} {\bibfnamefont {J.-C.}\ \bibnamefont {Besse}}, \bibinfo {author} {\bibfnamefont {M.}~\bibnamefont {Gabureac}}, \bibinfo {author} {\bibfnamefont {G.~J.}\ \bibnamefont {Norris}}, \bibinfo {author} {\bibfnamefont {A.}~\bibnamefont {Rosario}}, \bibinfo {author} {\bibfnamefont {F.}~\bibnamefont {Martin}}, \bibinfo {author} {\bibfnamefont {J.}~\bibnamefont
  {Martinez}}, \bibinfo {author} {\bibfnamefont {W.}~\bibnamefont {Amaya}}, \bibinfo {author} {\bibfnamefont {M.~W.}\ \bibnamefont {Mitchell}}, \bibinfo {author} {\bibfnamefont {C.}~\bibnamefont {Abellan}}, \bibinfo {author} {\bibfnamefont {J.-D.}\ \bibnamefont {Bancal}}, \bibinfo {author} {\bibfnamefont {N.}~\bibnamefont {Sangouard}}, \bibinfo {author} {\bibfnamefont {B.}~\bibnamefont {Royer}}, \bibinfo {author} {\bibfnamefont {A.}~\bibnamefont {Blais}},\ and\ \bibinfo {author} {\bibfnamefont {A.}~\bibnamefont {Wallraff}},\ }\bibfield  {title} {\bibinfo {title} {Loophole-free {Bell} inequality violation with superconducting circuits},\ }\href {https://doi.org/10.1038/s41586-023-05885-0} {\bibfield  {journal} {\bibinfo  {journal} {Nature}\ }\textbf {\bibinfo {volume} {617}},\ \bibinfo {pages} {265} (\bibinfo {year} {2023})}\BibitemShut {NoStop}%
\bibitem [{\citenamefont {Schaltegger}\ and\ \citenamefont {Balatsky}()}]{schaltegger2023}%
  \BibitemOpen
  \bibfield  {author} {\bibinfo {author} {\bibfnamefont {J.}~\bibnamefont {Schaltegger}}\ and\ \bibinfo {author} {\bibfnamefont {A.~V.}\ \bibnamefont {Balatsky}},\ }\bibfield  {title} {\bibinfo {title} {Gravitational phase shift in superconducting qubits},\ }\href@noop {} {\bibinfo  {journal} {(unpublished)}\ }\BibitemShut {NoStop}%
\bibitem [{\citenamefont {Bassi}\ \emph {et~al.}(2022)\citenamefont {Bassi}, \citenamefont {Cacciapuoti}, \citenamefont {Capozziello}, \citenamefont {Dell’Agnello}, \citenamefont {Diamanti}, \citenamefont {Giulini}, \citenamefont {Iess}, \citenamefont {Jetzer}, \citenamefont {Joshi}, \citenamefont {Landragin}, \citenamefont {Poncin-Lafitte}, \citenamefont {Rasel}, \citenamefont {Roura}, \citenamefont {Salomon},\ and\ \citenamefont {Ulbricht}}]{bassi2022}%
  \BibitemOpen
\bibfield  {journal} {  }\bibfield  {author} {\bibinfo {author} {\bibfnamefont {A.}~\bibnamefont {Bassi}}, \bibinfo {author} {\bibfnamefont {L.}~\bibnamefont {Cacciapuoti}}, \bibinfo {author} {\bibfnamefont {S.}~\bibnamefont {Capozziello}}, \bibinfo {author} {\bibfnamefont {S.}~\bibnamefont {Dell’Agnello}}, \bibinfo {author} {\bibfnamefont {E.}~\bibnamefont {Diamanti}}, \bibinfo {author} {\bibfnamefont {D.}~\bibnamefont {Giulini}}, \bibinfo {author} {\bibfnamefont {L.}~\bibnamefont {Iess}}, \bibinfo {author} {\bibfnamefont {P.}~\bibnamefont {Jetzer}}, \bibinfo {author} {\bibfnamefont {S.~K.}\ \bibnamefont {Joshi}}, \bibinfo {author} {\bibfnamefont {A.}~\bibnamefont {Landragin}}, \bibinfo {author} {\bibfnamefont {C.~L.}\ \bibnamefont {Poncin-Lafitte}}, \bibinfo {author} {\bibfnamefont {E.}~\bibnamefont {Rasel}}, \bibinfo {author} {\bibfnamefont {A.}~\bibnamefont {Roura}}, \bibinfo {author} {\bibfnamefont {C.}~\bibnamefont {Salomon}},\ and\ \bibinfo {author} {\bibfnamefont {H.}~\bibnamefont {Ulbricht}},\
  }\bibfield  {title} {\bibinfo {title} {A way forward for fundamental physics in space},\ }\href {https://doi.org/https://doi.org/10.1038/s41526-022-00229-0} {\bibfield  {journal} {\bibinfo  {journal} {npj Microgravity}\ }\textbf {\bibinfo {volume} {8}},\ \bibinfo {pages} {49} (\bibinfo {year} {2022})}\BibitemShut {NoStop}%
\bibitem [{\citenamefont {Chiao}\ \emph {et~al.}(2024)\citenamefont {Chiao}, \citenamefont {Inan}, \citenamefont {Scheibner}, \citenamefont {Sharping}, \citenamefont {Singleton},\ and\ \citenamefont {Tobar}}]{chiao2023}%
  \BibitemOpen
  \bibfield  {author} {\bibinfo {author} {\bibfnamefont {R.~Y.}\ \bibnamefont {Chiao}}, \bibinfo {author} {\bibfnamefont {N.~A.}\ \bibnamefont {Inan}}, \bibinfo {author} {\bibfnamefont {M.}~\bibnamefont {Scheibner}}, \bibinfo {author} {\bibfnamefont {J.}~\bibnamefont {Sharping}}, \bibinfo {author} {\bibfnamefont {D.~A.}\ \bibnamefont {Singleton}},\ and\ \bibinfo {author} {\bibfnamefont {M.~E.}\ \bibnamefont {Tobar}},\ }\bibfield  {title} {\bibinfo {title} {Gravitational {Aharonov-Bohm} effect},\ }\href {https://doi.org/10.1103/PhysRevD.109.064073} {\bibfield  {journal} {\bibinfo  {journal} {Phys. Rev. D}\ }\textbf {\bibinfo {volume} {109}},\ \bibinfo {pages} {064073} (\bibinfo {year} {2024})}\BibitemShut {NoStop}%
\bibitem [{\citenamefont {Toro\ifmmode~\check{s}\else \v{s}\fi{}}\ \emph {et~al.}(2022)\citenamefont {Toro\ifmmode~\check{s}\else \v{s}\fi{}}, \citenamefont {Cromb}, \citenamefont {Paternostro},\ and\ \citenamefont {Faccio}}]{toros2022}%
  \BibitemOpen
  \bibfield  {author} {\bibinfo {author} {\bibfnamefont {M.}~\bibnamefont {Toro\ifmmode~\check{s}\else \v{s}\fi{}}}, \bibinfo {author} {\bibfnamefont {M.}~\bibnamefont {Cromb}}, \bibinfo {author} {\bibfnamefont {M.}~\bibnamefont {Paternostro}},\ and\ \bibinfo {author} {\bibfnamefont {D.}~\bibnamefont {Faccio}},\ }\bibfield  {title} {\bibinfo {title} {Generation of entanglement from mechanical rotation},\ }\href {https://doi.org/10.1103/PhysRevLett.129.260401} {\bibfield  {journal} {\bibinfo  {journal} {Phys. Rev. Lett.}\ }\textbf {\bibinfo {volume} {129}},\ \bibinfo {pages} {260401} (\bibinfo {year} {2022})}\BibitemShut {NoStop}%
\bibitem [{\citenamefont {Hehl}\ \emph {et~al.}(1976)\citenamefont {Hehl}, \citenamefont {von~der Heyde}, \citenamefont {Kerlick},\ and\ \citenamefont {Nester}}]{RevModPhys.48.393}%
  \BibitemOpen
  \bibfield  {author} {\bibinfo {author} {\bibfnamefont {F.~W.}\ \bibnamefont {Hehl}}, \bibinfo {author} {\bibfnamefont {P.}~\bibnamefont {von~der Heyde}}, \bibinfo {author} {\bibfnamefont {G.~D.}\ \bibnamefont {Kerlick}},\ and\ \bibinfo {author} {\bibfnamefont {J.~M.}\ \bibnamefont {Nester}},\ }\bibfield  {title} {\bibinfo {title} {General relativity with spin and torsion: Foundations and prospects},\ }\href {https://doi.org/10.1103/RevModPhys.48.393} {\bibfield  {journal} {\bibinfo  {journal} {Rev. Mod. Phys.}\ }\textbf {\bibinfo {volume} {48}},\ \bibinfo {pages} {393} (\bibinfo {year} {1976})}\BibitemShut {NoStop}%
\bibitem [{\citenamefont {Bakke}\ \emph {et~al.}(2009)\citenamefont {Bakke}, \citenamefont {Furtado},\ and\ \citenamefont {Nascimento}}]{bakke2009}%
  \BibitemOpen
  \bibfield  {author} {\bibinfo {author} {\bibfnamefont {K.}~\bibnamefont {Bakke}}, \bibinfo {author} {\bibfnamefont {C.}~\bibnamefont {Furtado}},\ and\ \bibinfo {author} {\bibfnamefont {J.}~\bibnamefont {Nascimento}},\ }\bibfield  {title} {\bibinfo {title} {Gravitational geometric phase in the presence of torsion},\ }\href {https://doi.org/10.1140/epjc/s10052-009-0944-z} {\bibfield  {journal} {\bibinfo  {journal} {Eur. Phys. J. C}\ }\textbf {\bibinfo {volume} {60}},\ \bibinfo {pages} {501} (\bibinfo {year} {2009})}\BibitemShut {NoStop}%
\bibitem [{\citenamefont {Crowder}\ and\ \citenamefont {Lanzagorta}(2023)}]{crowder2023}%
  \BibitemOpen
  \bibfield  {author} {\bibinfo {author} {\bibfnamefont {T.}~\bibnamefont {Crowder}}\ and\ \bibinfo {author} {\bibfnamefont {M.}~\bibnamefont {Lanzagorta}},\ }\bibfield  {title} {\bibinfo {title} {Gravitationally invariant subspaces in quantum computing},\ }\href {https://doi.org/10.1007/s11047-022-09938-7} {\bibfield  {journal} {\bibinfo  {journal} {Nat. Comput.}\ } (\bibinfo {year} {2023})}\BibitemShut {NoStop}%
\end{thebibliography}%

%%%%%%%%%%%%%%%%%%%%%%%%%%%%%%%

\end{document}